\documentclass[aps,preprint,showpacs]{revtex4}
\usepackage{graphicx}
\newcommand{\TT}{{\tilde T}}
\newcommand{\nT}{{\tilde n}}
\newcommand{\mT}{{\tilde m}}
\newcommand{\muT}{{\tilde \mu}}
\newcommand{\ET}{{\tilde E}}
\newcommand{\gT}{{\tilde g}}
\newcommand{\gaT}{{\tilde \alpha}}
\newcommand{\gamT}{{\tilde \gamma}}
\begin{document}
\title{Molecular formations in ultracold mixtures of interacting and noninteracting atomic gases}
%
%
\author{T.~Nishimura\footnote{nimut@tmu.ac.jp} and
        A.~Matsumoto}
\affiliation{Department of Physics, Tokyo Metropolitan University, 
                  1-1 Minami-Ohsawa, Hachioji, Tokyo 192-0397, Japan} 
\author{H. Yabu\footnote{yabu@se.ritsumei.ac.jp}}
\affiliation{Department of Physics, Ritsumeikan University, Kusatsu, Shiga  
             525-8577, Japan} 
\begin{abstract}
Atom-molecule equilibrium for molecular formation processes 
is discussed 
for boson-fermion, fermion-fermion, and boson-boson mixtures
of ultracold atomic gases
in the framework of quasichemical equilibrium theory.
After presentation of the general formulation,
zero-temperature phase diagrams of the atom-molecule equilibrium states 
are calculated analytically; 
molecular, mixed, and dissociated phases are shown to appear 
for the change of the binding energy of the molecules. 
The temperature dependences of 
the atom or molecule densities are calculated numerically, 
and finite-temperature phase structures are obtained
of the atom-molecule equilibrium in the mixtures.
The transition temperatures 
of the atom or molecule Bose-Einstein condensations 
are also evaluated from these results. 
Quantum-statistical deviations of the law of mass action
in atom-molecule equilibrium, 
which should be satisfied 
in mixtures of classical Maxwell-Boltzmann gases, 
are calculated, 
and the difference in the different types of quantum-statistical effects is clarified. 
Mean-field calculations with interparticle interactions 
(atom-atom, atom-molecule, and molecule-molecule) 
are formulated, where
interaction effects are found to give 
the linear density-dependent term 
in the effective molecular binding energies.
This method is applied to calculations 
of zero-temperature phase diagrams, 
where new phases with coexisting local-equilibrium states 
are shown to appear
in the case of strongly repulsive interactions.
\end{abstract} 
\maketitle
%
%
\section{Introduction}
%
%

The experimental success of the Bose-Einstein condensation (BEC) \cite{nobel}
of the trapped ultracold atomic gas has made 
much progress 
in the physics of quantum gases\cite{Dalfovo,Pethik,Pitaevskii}, 
which includes Fermi-degenerate systems
and Bose-Fermi mixtures.

Recently, using the Feshbach-resonance method,
molecular formations have been performed experimentally 
in ultracold atomic gases 
for two fermions\cite{Regal,Cubizolles,Jochim} and 
two bosons\cite{Wynar,Jochim2};
Bose-Einstein condensations  
have been observed for the thus created molecules 
of two fermions\cite{Greiner1,Zwierlein} and two bosons\cite{Jochim2}.
In these experiments, 
the molecular binding energies  
can be tuned by continuous changes of the applied magnetic fields;
especially, bound molecular states can be changed into resonances 
by shifting the binding energies above the atom-atom scattering thresholds.

One of the interesting applications 
of ultracold molecules 
is the observation of the BEC-BCS crossover 
in experiments with atomic Fermi gases: 
continuous crossover between the strong-coupling molecular BEC 
and weak-coupling BCS superconducting states
\cite{SBB1,Leggett,Nozieres,Randeira,SRS1,Ropk1}.
In ultracold atomic-gas experiments, 
the crossover occurred
by a change of the molecular binding energy 
through the Feshbach resonance method\cite{Regal2,Bartenstein,Zwierlein2}, 
and the experimental success has led to
a lot of experimental and theoretical works 
on the crossover\cite{Ohashi,Falco,Chen} 
and molecular BEC physics\cite{Chin,Nygaard1}.  

Molecules in optical lattice potentials 
are also an interesting problem. 
After the first observation of molecular formations
using ${}^{87}$Rb\cite{Rom},
many experiments have been performed 
on the long-lived state of the ${}^{87}$Rb molecule\cite{Thalhammer}
, ${}^{40}$K difermion\cite{Koel}, 
and ${}^{87}$Rb-${}^{40}$K boson-fermion 
heteronuclear molecules\cite{Ospelkaus}.
Quantum degeneracy has also been observed 
in gases of ${}^{87}$Rb molecules\cite{Ryu}.
The phase structure of the lattice-trapped atomic gas, 
which is produced through interparticle correlations, 
has much interest in relation to the strongly correlated 
condensed-matter system described by the Hubbard model. 
Thus many experimental and theoretical works 
have been done: for example, 
on the superfluid-Mott insulator transition 
in bosonic\cite{Greiner2,Mandel,Bartouni,Jaksch,Jack}, 
fermionic \cite{Roati,Fujiwara}, 
and boson-fermion systems\cite{Miyakawa1}.

Other theoretical studies on ultracold molecules
include coherent molecular solitons\cite{Drummond}, 
coherent photoassociation\cite{Javanainen}, 
and coherent intercondensate exchange 
between atoms and molecules\cite{Timmermans}.

In this paper, 
we discuss atom-molecule equilibrium 
for boson-fermion, fermion-fermion, and boson-boson mixtures 
of ultracold atomic gases 
with quasichemical equilibrium theory, 
an extension of classical chemical equilibrium theory\cite{Atkins,Laidler}
for the quantum many-body problem,
which was originally developed for the electron system 
in superconductors\cite{SBB1,Leggett}. 
Our special interest is in applications of this method 
to boson-fermion and boson-boson mixtures, 
especially occurrences of atom or molecule BEC, 
as shown in some works 
for boson-fermion mixtures\cite{Yabu1,Yabu2,Strozhenko,Nygaard2}.
In addition, in contrast to the fermionic system,  
where many-body quantum calculations based on the microscopic model 
have been established, 
the singularities from the boson degrees of freedom 
sometimes cause problems in calculations  
of boson-fermion and boson-boson mixtures;
quasichemical equilibrium theory can give definite solutions 
in such cases.
As another interest of this approach, 
it should be pointed out 
that we can easily include interparticle interactions, 
especially atom-molecule and molecule-molecule ones,
which are sometimes omitted in many-body calculations.
The quasichemical approach gives the equilibrium structures 
in a less model-independent way. 
It turned out that the effects of these interactions should change 
the atom-molecule equilibrium structures drastically 
in the strong-coupling region.

In Sec.~II, a general formulation of quasichemical equilibrium theory 
is presented
for molecular formation or dissociation processes 
in noninteracting atomic-gas mixtures, 
and in Sec.~III, 
the method is applied to boson-fermion, fermion-fermion, 
and boson-boson mixtures 
and the atom-molecule equilibrium structures 
are shown at zero and finite temperatures. 
Special attention is paid to the condition 
on the occurrences of atom or molecule BEC;
the shifts of the BEC transition temperatures are discussed 
from the molecular binding energy effects.
In Sec.~IV, 
the law of mass action, 
which is satisfied in chemical processes 
with classical Boltzmann statistics,
is shown to deviate in ultracold molecular formation or dissociation processes 
by quantum-statistical effects (the law of quantum mass action). 
In Sec.~V,
we extend the quasichemical theory to include interparticle interactions 
for $s$-wave scattering processes
(three kinds of atom-atom ones, two kinds of atom-molecule ones, and one kind of molecule-molecule one 
in combination)
in molecular formation or dissociation processes.
In the mean-field approximation, 
the original six coupling constants of the interactions 
are shown to integrate into two parameters.
It allows discussions of the interaction effects to be very clear.
The formulations are applied to molecular formation or dissociation processes 
in interacting mixtures, 
and we discuss the change of the equilibrium structures at $T=0$
through interaction effects; 
new phases with coexisting local-equilibrium states 
are shown to appear in the case of strongly repulsive interactions.
Section VI is devoted to a summary and outlook. 

We should comment that a relativistic extension is also possible
of quasichemical theory and applications to diquark condensates 
in quark matter have been done in \cite{Nawa}. 

%
\section{Quasichemical Equilibrium Theory on Molecular Formation 
         in Atomic-gas mixtures}
%
\subsection{Molecular formation or dissociation process}

Let us take an atomic-gas mixture consisting of 
two atomic spices $A_1$ and $A_2$ 
with masses $m_1$ and $m_2$; 
in quantum statistics, 
$A_{1,2}$ are bosons or fermions, 
so that we have three kinds of combinations:
boson-boson (BB), fermion-fermion (FF), and boson-fermion (BF).

To develop a quasichemical equilibrium theory, 
we consider a molecular formation or dissociation process
in the mixture:
\begin{equation}
     A_1 +A_2 \longleftrightarrow (A_1 A_2)=M,  
\label{EqB1}  
\end{equation}
where $M$ is a composite molecule 
with mass $m_M$, which is bosonic in BB or FF mixtures 
and fermionic in BF mixtures. 

The mass defect of the molecule is defined as
$\Delta{m_M} \equiv (m_M-m_1-m_2)$.
The boundmolecule ($\Delta{m_M} < 0$) 
is stable in both vacuum and gases, 
and has molecular binding energy $\Delta{E} =\Delta{m_M} c^2$, 
where $c$ is the velocity of light in vacuum. 
In contrast, 
the resonance ($\Delta{m_M} > 0$) 
is unstable at least in vacuum; 
however, 
resonance states can exist stably in gases, 
so that we consider both bound-molecule 
and resonance cases. 

Here we take the quasiparticle picture 
wherein the system consists of atoms $A_1$ and $A_2$ 
and molecule $M$, which are quasiparticles\cite{Landau}.
In this picture, 
two-body interactions between ``bare'' atoms 
bring about two-body correlations in the mixture; 
their major effects are the creation of the composite molecule $M$ 
with binding energy $\Delta{E_M}$ 
and  
atoms and molecules, 
which are quasiparticle in the mixture,
and interact through residual interactions, 
which are generally different from the original interaction 
between bare atoms.
These quasiparticle interactions are generally regarded to be weak, 
and we neglect them in the first part of this paper. 
In Sec.~IV, we introduce the quasiparticle interactions
and discuss their effects on atom-molecule equilibrium 
within the mean-field approximation.

The equilibrium condition for the process (\ref{EqB1}) 
is given by 
\begin{equation}
     \mu_1 +\mu_2 -\mu_M =\Delta{E_M},
\label{EqB2}  
\end{equation}
where $\mu_{1,2,M}$ are chemical potentials 
of $A_1$, $A_2$, and $M$. 

The molecular binding energy $\Delta{E_M}$ in (\ref{EqB2}) is generally very small 
($\sim 10^{-(5 \sim 10)}\,{\rm eV}$) 
in molecular formation or dissociation processes in ultracold atomic-gases, 
and it takes the same order of magnitude 
with the chemical potentials of atoms or molecules $\mu_\alpha$
at ultralow temperature ($T ={\rm \mu K} - {\rm nK}$). 
Thus, the term $\Delta{E_M}$ cannot be omitted in (\ref{EqB2}). 

For free uniform gases, 
the particle densities are given by the Bose and Fermi statistics:
\begin{eqnarray}
     n_\alpha &=& \frac{1}{(2\pi)^3} 
             \int\frac{d^3{\bf k}}{
                   e^{(\varepsilon_\alpha -\mu_\alpha)/k_B T} -1} + n_\alpha^{(0)}
         \equiv \frac{1}{(\lambda_{T,\alpha})^3}
                B_{3/2}(-\mu_\alpha/k_B T) + n_\alpha^{(0)} 
      \hbox{(for boson $A_\alpha$)}, \label{EqB6} \\ 
     n_\alpha &=& \frac{1}{(2\pi)^3} 
             \int\frac{d^3{\bf k}}{
                   e^{(\varepsilon_\alpha -\mu_\alpha)/k_B T} +1} 
         \equiv \frac{1}{(\lambda_{T,\alpha})^3}
                F_{3/2}(-\mu_\alpha/k_B T) 
     \hbox{(for fermion $A_\alpha$)}, \label{EqB7} 
\end{eqnarray}
where $k_B$ is the Boltzmann constant and $\lambda_{T,\alpha}$ 
is the thermal de~Broglie wave length of particle $A_\alpha$
at temperature $T$:
\begin{equation}
     \lambda_{T,\alpha} =\sqrt{\frac{2\pi\hbar^2}{m_\alpha k_B T}}. 
\label{EqB12}
\end{equation} 
The one-particle energy $\varepsilon_\alpha$ in (\ref{EqB6}) and (\ref{EqB7}) is give by 
$\varepsilon_\alpha = \hbar^2 {\bf k}^2 / (2 m_\alpha)$, 
where ${\bf k}$ is the wave-number vector of particle $A_\alpha$.

The $B_A$ and $F_A$ in (\ref{EqB6}) and (\ref{EqB7}) are the Bose and Fermi functions:
\begin{eqnarray}
     B_A(\nu) &=& \frac{1}{\Gamma(A)}
                  \int_0^\infty
                  \frac{x^{A-1} dx}{e^{x+\nu}-1}, \label{EqB10}\\
     F_A(\nu) &=& \frac{1}{\Gamma(A)}
                  \int_0^\infty
                  \frac{x^{A-1} dx}{e^{x+\nu}+1}, \label{EqB11}
\end{eqnarray}
where $\nu$ corresponds to the fugacity and 
$\Gamma(A)$ is the gamma function.
The Bose function $B_A$ can be written
with the Appel function $\phi(z,s)$ as $B_A(\nu) =\phi(A,e^\nu)$, 
and the Fermi function is expressed with the Bose functions 
$F_A(\nu) =B_A(\nu) -2^{1-A} B_A(2\nu)$. 

The Bose function $B_A$ in (\ref{EqB10}) converges 
in $\nu \geq 0$ (or $\mu_\alpha \leq 0$ in the chemical potential) 
and becomes singular 
at $\mu_\alpha=0$, 
with which a phase transition occurs to the Bose-Einstein condensate
of the boson $A_\alpha$.
In the BEC region, 
the chemical potential vanishes 
and the condensed density $n_\alpha^{(0)}$ in (\ref{EqB6}) 
takes a finite value 
(it vanishes in the normal region). 

In the process (\ref{EqB1}), 
the particle-number conservations for $A_{1,2}$ give the constraints
\begin{equation}
     n_1 +n_M =n_{1,t},  \qquad
     n_2 +n_M =n_{2,t},  
\label{EqB14}  
\end{equation}
where $n_{1,t}$ and $n_{2,t}$ are the total number densities 
of atoms $A_{1,2}$, 
which consist of isolated atoms and constituents in
the composite molecule $M$.

Solving Eqs. (\ref{EqB2}) and (\ref{EqB14}) 
with (\ref{EqB6}) and (\ref{EqB7}), 
we can determine the densities $n_{1,2,M}$ in equilibrium at temperature $T$
for the parameters $m_\alpha$, $n_{\alpha,t}$ ($\alpha=1,2$), 
and $\Delta{E}$.

\subsection{Scaled variables}

We now introduce scaled dimensionless variables; 
they simplify the form of the above equations 
and greatly reduce the number of parameters.

The scaled masses $\mT_\alpha$ are defined by 
$\mT_\alpha=m_\alpha/m_M$, 
where $m_M =m_1 +m_2 +\Delta{m_M}$. 
In the ultracold atomic-gas system, 
the mass defect ($\Delta{m_M} \sim 10^{-5}-10^{-10}\,{\rm eV}$) 
is highly smaller than the atom or molecule masses ($\sim{\rm GeV}$);
thus, we use the approximation $m_M \sim m_1+m_2$ 
(the conservation law of mass in chemical processes) 
throughout this paper. 

We introduce $n_t \equiv n_{1,t}+n_{2,t}$ 
and $E_s \equiv \hbar^2 (n_t)^{2/3}/m_M$ 
as scaling quantities for
the particle-number densities and 
the energies, respectively: 
for example, 
$\nT_\alpha =n_\alpha/n_t$ ($\alpha=1$, $2$, and $M$), 
$\TT = k_B T / E_s$, 
$\Delta{\tilde E}_M =\Delta{E}_M / E_s$, etc. 
$E_s$ can be interpreted as having the meaning of the Fermi energy 
for the fermionic matter of fermions with mass $m_M$
with density $n_t$, 
but we use it simply as a scaling quantity with the dimension of energy.

Using the scaled quantities, 
the equilibrium conditions (\ref{EqB2}) and (\ref{EqB14}) become 
\begin{equation}
     \muT_1 +\muT_2 -\muT_M 
          =\Delta{\tilde E}_M,
\label{EqB17}
\end{equation}
\begin{equation}
     \nT_1 +\nT_M =\nT_{1,t}, \qquad 
     \nT_2 +\nT_M =\nT_{2,t}.  
\label{EqB18a}
\end{equation}
The Bose and Fermi statistics formulas (\ref{EqB6}) and (\ref{EqB7}) 
become
\begin{eqnarray}
     \nT_\alpha &=& \left[ \frac{\mT_\alpha \TT}{2\pi} 
                           \right]^{3/2}
                           B_{3/2}(-\muT_\alpha/\TT) + \nT_\alpha^{(0)} 
     \qquad \hbox{(for boson $A_\alpha$),} \label{EqB19} \\ 
     \nT_\alpha &=& \left[ \frac{\mT_\alpha \TT}{2\pi} 
                           \right]^{3/2}
                           F_{3/2}(-\muT_\alpha/\TT) 
     \qquad \hbox{(for fermion $A_\alpha$).} \label{EqB20} 
\end{eqnarray}

Now the problem of obtaining the atom-molecule equilibrium states is reduced 
to solving Eqs.~(\ref{EqB17}) and (\ref{EqB18a}) 
with (\ref{EqB19}) and (\ref{EqB20}) 
for temperature $\TT$,
molecular binding energy $\Delta{\tilde E}_M$, 
and mass and total density of $A_1$ ($\mT_1$, $\nT_{1,t}$).
The scaled atomic masses and the scaled total densities 
take the values of $0 \leq \mT_\alpha, \nT_{\alpha,t} \leq 1$ 
and satisfy
\begin{equation}
     \mT_1 +\mT_2 =1,  \qquad
     \nT_{1,t} +\nT_{2,t} =1. 
\label{EqB15}
\end{equation}
These constraints play 
the role of reducing the number of independent parameters.

%
\section{Molecular Formations in Noninteracting Atomic-gas Mixtures}
%

%
%
\subsection{Bose-Einstein condensation and Fermi degeneracy}

Before presenting the calculational results, 
we discuss two interesting quantum effects: 
BEC for bosons 
and Fermi degeneracy (FD) for fermions.

As discussed in the previous section, 
the phase transition to the BEC of the boson $A_\alpha$ occurs 
when $\muT_\alpha =0$ in (\ref{EqB19}); 
the transition temperature $T_C$ is
\begin{equation}
     \TT_C =\frac{2\pi (\nT_\alpha)^{2/3}}{\mT_\alpha} 
                   [\zeta(3/2)]^{-2/3}
           \sim 3.313 \frac{(\nT_\alpha)^{2/3}}{\mT_\alpha},  
\label{EqC1}
\end{equation}
where $\zeta(3/2)$ is the Riemann zeta function.
The condensed and thermal parts of 
the number density at $T < T_C$ are given by (\ref{EqB19});
with scaled quantities, they become
\begin{equation}
     \nT_\alpha^{(th)} 
          =\left( \frac{ \mT_\alpha \TT}{2\pi} \right)^{3/2} 
           \zeta(3/2),   \qquad
     \nT_\alpha^{(0)} =\left[ 1 
                             -\left( \frac{\TT}{\TT_C} \right)^{3/2}
                       \right] \nT_\alpha. 
\label{EqC2}
\end{equation}
From these equations, 
we can find that, 
if bosons $A_\alpha$ exist in the mixture at $\TT =0$ 
($\nT_\alpha \neq 0$), 
all $A_\alpha$ should condense into the BEC 
($\nT_\alpha^{(0)} =\nT_\alpha$ at $\TT =0$).

Different from bosons,
fermions go into the FD state at very low temperatures;
especially, at $\TT=0$, Eq.~(\ref{EqB20}) becomes
\begin{equation}
     \nT_\alpha =\frac{\sqrt{2}(\mT_\alpha)^{3/2}}{3\pi^2}
                        (\muT_{F,\alpha})^{3/2}, 
\label{EqC4}
\end{equation}
where $\mu_{F,\alpha} \equiv \muT_\alpha(\TT=0)$ 
is the Fermi energy. 

The transition into the FD state 
is not a phase transition, but a continuous one, 
so that, different from the BEC $\TT_C$ in (\ref{EqC1}), 
no clear boundaries exist for FD. 
Instead, as an estimation of the occurrence of FD, 
we use the temperature 
at $\muT_{F,\alpha} =0$:
\begin{equation}
     \TT_{F,\alpha} =\frac{( 6\pi^2 \nT_{F,\alpha})^{2/3}}{2\mT_{F,\alpha}}
                     \sim 7.569 \frac{ (\nT_\alpha)^{2/3} }{\mT_\alpha}.
\label{EqC6}
\end{equation}

Because of the permitted ranges of the scaled densities 
$0 \leq \nT \leq 1$, 
we can find $\TT_C \sim \TT_F$ from Eqs.~ (\ref{EqC1}) and (\ref{EqC6}) 
as a rough estimation; 
quantum effects begin to appear 
at the same order of temperature in both noninteracting Bose and Fermi gases.

\subsection{Molecular formations in the BF mixture}

Let us consider the BF mixture ($A_1 = B$, $A_2 = F$) 
with the molecular formation process 
$B + F \leftrightarrow M = (BF)$. 
The atom or molecule densities $\nT_{B,F,M}$ in atom-molecule equilibrium 
are obtained
from the equilibrium condition (\ref{EqB17}) 
with Eqs. (\ref{EqB18a})-(\ref{EqB20}).
Interparticle interaction effects are neglected, 
which will be discussed in a later section.

To understand the overall structure of the atom-molecule equilibrium, 
we consider the phase diagram at $T=0$, 
which is obtained with analytical calculations 
as shown in Appendix B.
Fig.~\ref{fig1} shows the $T=0$ phase diagram 
of the BF mixture with the same atom masses $\mT_B =\mT_F =1/2$ 
for the molecular binding energy ($\Delta\ET_M$) 
and a total density of $B$ ($\nT_{B,t}$). 
Because of the constraints (\ref{EqB15}), 
$\Delta\ET_M$ and $\nT_{B,t}$ are unique parameters
to determine the equilibrium state.
The bracketed letters in the regions or lines in Fig.~\ref{fig1} represent 
the species which exist in equilibrium at $T=0$; 
for example, $(B,F,M)$ in the central region 
means that atoms $B$ and $F$ and molecule $M$ 
coexist there, etc. 

\begin{figure}[ht]
  \begin{center}
       \includegraphics[scale=0.3,angle=-90]{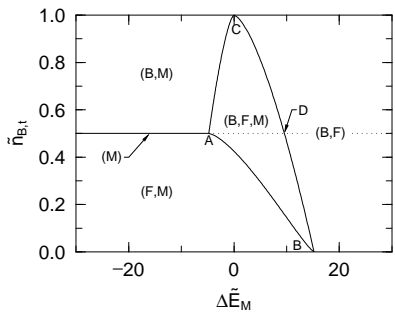}
  \end{center}
\caption{
The $T=0$ phase diagram of BF mixtures with the same atom masses  
in the $\Delta{\ET_M}$-$\nT_{B,t}$ plane.
The bracketed letters show 
what kinds of particles exist in equilibrium at $T=0$.
}
\label{fig1}
\end{figure}

From Fig.~\ref{fig1}, we find that the BF mixture at $T=0$ shows 
the structure that, in the area left of the border $BAC$ 
($\Delta\ET_M \lesssim 0$), 
the states with molecules as many as possible 
become stable (molecular states)
and, in the area right of the boundary $BC$, 
all molecules dissociate into atoms $B$ and $F$ (dissociated states).
Between these two areas, 
mixed states of atoms and molecules become stable 
in the sense of equilibrium.

In Fig.~\ref{fig1}, we can read off the existence condition
of the BEC of atom $B$; 
it is in the regions 
with the bracketed $B$ 
because the bosons always condense into a BEC at $T=0$.

\begin{figure}[ht]
  \begin{center}
    \begin{tabular}{cc}
       \includegraphics[scale=0.3,angle=-90]{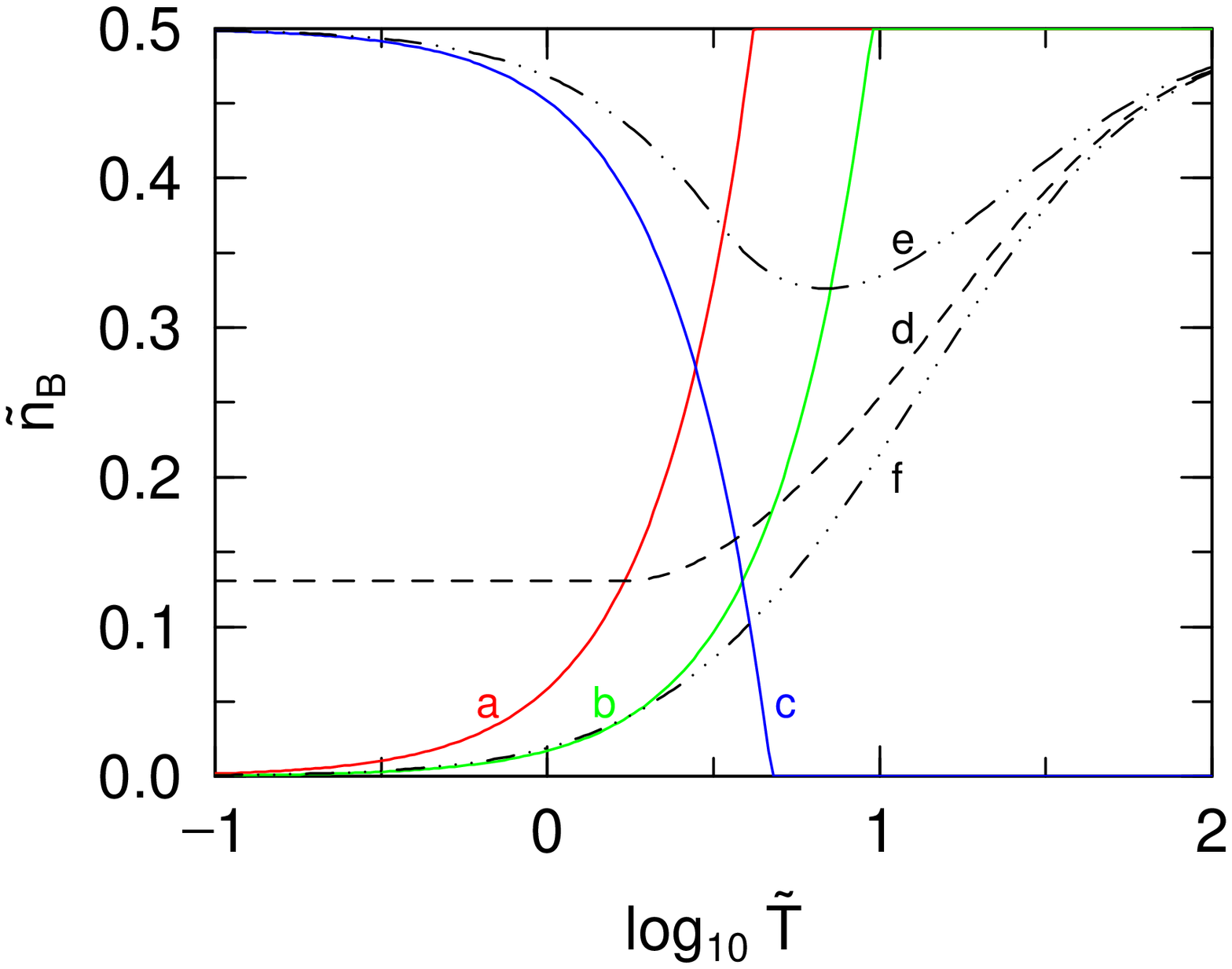} &
       \includegraphics[scale=0.3,angle=-90]{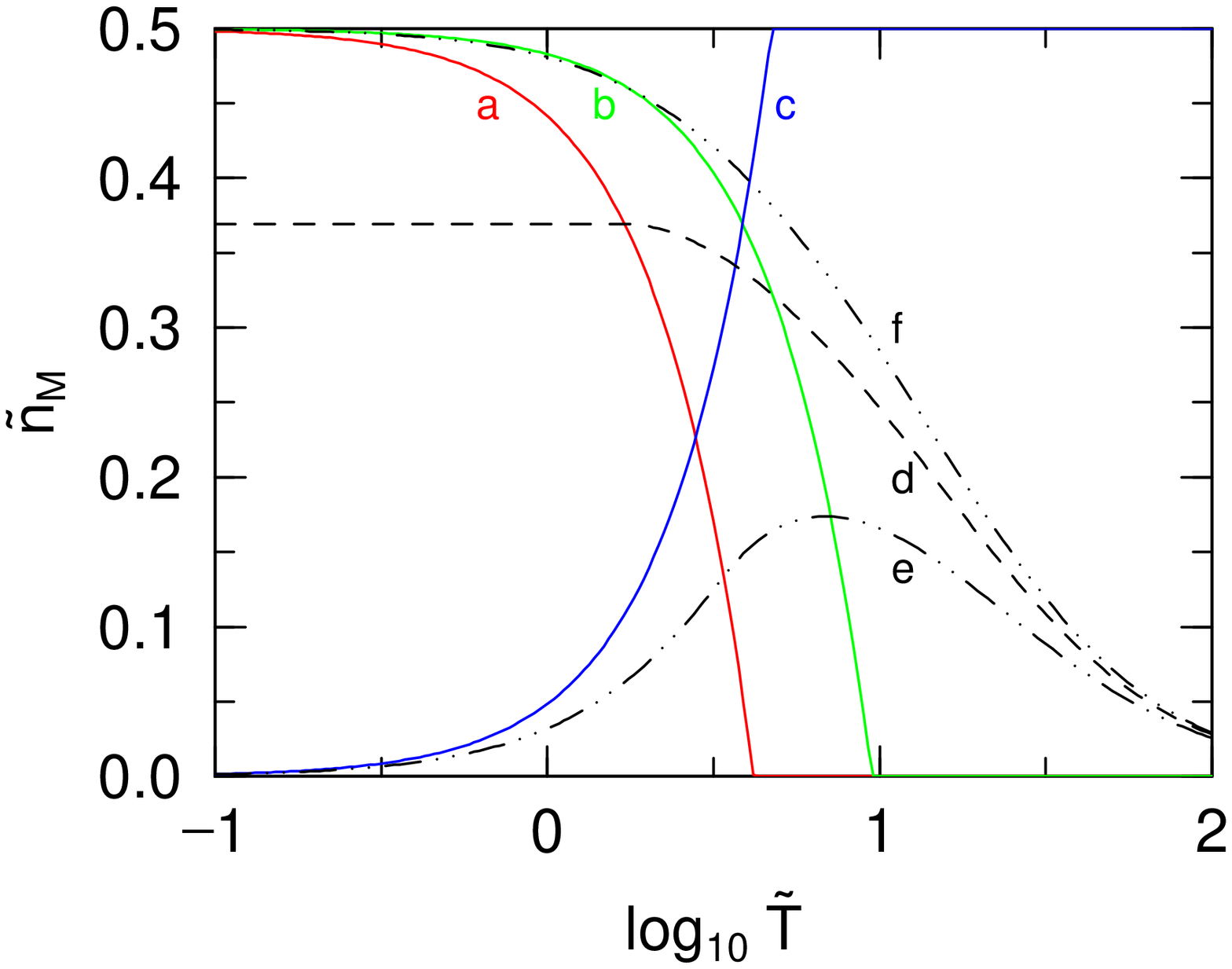}
    \end{tabular}
  \end{center}
\caption{(Color online) 
Temperature dependences 
of the number densities $\nT_B$ of the bosonic atom $B$ [(a), left] 
and $\nT_M$ of the molecule $M$ [(b), right] 
in the BF mixture 
with the same atomic masses and the same total atomic densities.
The lines $d$ (dashed line), $e$ (double-dot-dashed line) and $f$ (triple-dot-dashed line) 
are for mixtures with 
$\Delta{\ET_M} =0$, $9.57$ (resonance state), and $-4.78$
(bound state), respectively. 
The BEC $T_C$ line of $B$ (line $a$) 
and the $\muT_F=0$ lines 
for $F$ (line $b$) 
and $M$ (line $c$) are also drawn. }
\label{fig2}
\end{figure}

We now turn to the atom-molecule equilibrium 
of the BF mixture at finite temperatures, 
which are obtained in numerical calculations 
using (\ref{EqB7}) and (\ref{EqB17}).
Fig.~\ref{fig2} shows the temperature dependences of 
the scaled densities $\nT_B$ and $\nT_M$ 
in the BF mixture with masses 
$\mT_B =\mT_F =1/2$ 
and the same total atomic-number densities $\nT_{B,t} =\nT_{F,t} =1/2$.
The lines $d$, $e$, and $f$ in Fig.~\ref{fig2} are 
for several values of the molecular binding energies: 
$\Delta{\tilde E}_M =0$, $-4.78$, and $9.57$, respectively. 
The BEC line (\ref{EqC1}) and the $\muT_F =0$ line, Eq.~(\ref{EqC6}), 
are also drawn in Fig.~\ref{fig2} as lines $a$ and $b$, respectively; 
they show the boundaries
where quantum-statistical effects begin to occur. 

Fig.~\ref{fig2} shows that, 
at the high-temperature limit, 
$\nT_M$ converge to $0$ 
(complete dissociation of molecules into atoms). 
It can be explained by the energy-entropy balance in the free energy. 
The atom-molecule mixed states 
have a reduction in the energy part of the free energy; 
in contrast, the dissociated states 
have a reduction of the entropy part 
because the state density of $B+F$ is larger than that of molecules. 
With increasing temperature, 
the entropy contribution becomes large 
and the reaction process swings to the dissociation of molecules.
This mechanism can also be applied to BB or FF mixtures 
because the binding energy and quantum-statistical properties 
are less effective at high temperature. 

The equilibrium states  
have a large dependence on the binding energy $\Delta\ET_M$
around and below the BEC $T_C$ and $\muT_F=0$ lines.
At $T=0$, 
they converge to states belonging to the phases 
on the horizontal $\nT_{F,t}=1/2$ line in Fig.~\ref{fig1}: 
the molecular, mixed and dissociated phases 
divided by the points,
$\Delta\ET_M = \Delta_1 \equiv -(3\pi^2)^{2/3}/2 \approx -4.78$ (point A) 
and $\Delta_2 \equiv (3\pi^2)^{2/3} \approx 9.57$ (point D). 
The temperature dependence in Fig.~\ref{fig2} is also 
classified with the values of $\Delta\ET_M$: 
(i) $\Delta\ET_M \le \Delta_1$ 
(lines $f$). 
With decreasing $T$, 
the $\nT_B$ becomes small and goes
below the BEC $T_C$ line 
(no BEC of atom $B$ for all temperatures). 
(ii) $\Delta_1 < \Delta\ET_M < \Delta_2$
(line $d$).
With decreasing $T$, 
$\nT_B$ (and also $\nT_F$) and $\nT_M$ 
run into the BEC and $\muT \ge 0$ regions, 
and converge into finite values. 
(iii) $\Delta_2 \le \Delta\ET_M$
(line $e$).
With decreasing $T$, 
$\nT_M$ takes the maximum value 
on the right of the $\muT =0$ line (line $c$)
and decreases to $\nT_M=0$ along the line from below. 

The atom-molecule equilibrium 
in the BF mixtures 
can be explained 
by competition between the quantum-statistical effects
and the binding energy of $M$; 
the molecule states give a free-energy reduction
because of $\Delta\ET_M <0$, 
but at low $T$, 
the molecules constitute the FD states,
which have large kinetic energies 
coming from the occupied high-energy one-particle states.
In contrast, in dissociated states,
in spite of the kinetic-energy contribution
from the FD states of the fermions $F$, 
the bosons $B$ can reduce the kinetic energy largely 
as they condense into the BEC at low-$T$.
Thus, depending on the positive or negative value 
of $\Delta\ET_M$, 
the dissociated or molecular state becomes stable, 
and a mixed phase appears between them. 

\subsection{Molecular formations in the FF mixture}

In the FF mixture ($A_1 =F1$, $A_2 =F2$), 
we consider atom-molecule equilibrium 
through the process
$F1 +F2 \leftrightarrow M =(F1F2)$ 
in the same way as in the BF mixture. 
In this section, 
we take the FF mixture with the same atomic masses 
$\mT_{F1} =\mT_{F1} =1/2$. 

\begin{figure}[ht]
  \begin{center}
       \includegraphics[scale=0.3,angle=-90]{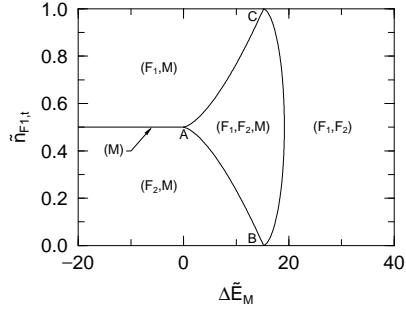}
  \end{center}
\caption{
The $T=0$ phase diagram of the FF mixture 
with the same atom masses 
in the $\Delta{\ET_M}$-$\nT_{F1,t}$ plane.
The bracketed letters show 
what kinds of particles exist in equilibrium at $T=0$.
}
\label{fig4}
\end{figure}

Fig.~\ref{fig4} shows 
the $T=0$ phase diagram of the FF mixture. 
The notation for each phase 
is the same as that of the BF mixture (see Fig.~\ref{fig1}). 
Analytical expressions of the phase-boundary lines and points 
in Fig.~\ref{fig4} are given in Appendix B. 

The topological structure of the equilibrium phases in Fig.~\ref{fig4} is 
the same as that in the BF mixture (Fig.~\ref{fig1}); 
it consists of 
a ``molecular phase''
to the left of the border $BAC$,
a ``dissociated phase'' 
to the right of the boundary $BC$, 
and 
a ``mixed phase''
between them. 

The existence condition
of BEC of the bosonic molecules $M$ 
is also read off in Fig.~\ref{fig4}
as the regions with the symbol $M$ in brackets.

\begin{figure}[ht]
  \begin{center}
    \begin{tabular}{cc}
       \includegraphics[scale=0.3,angle=-90]{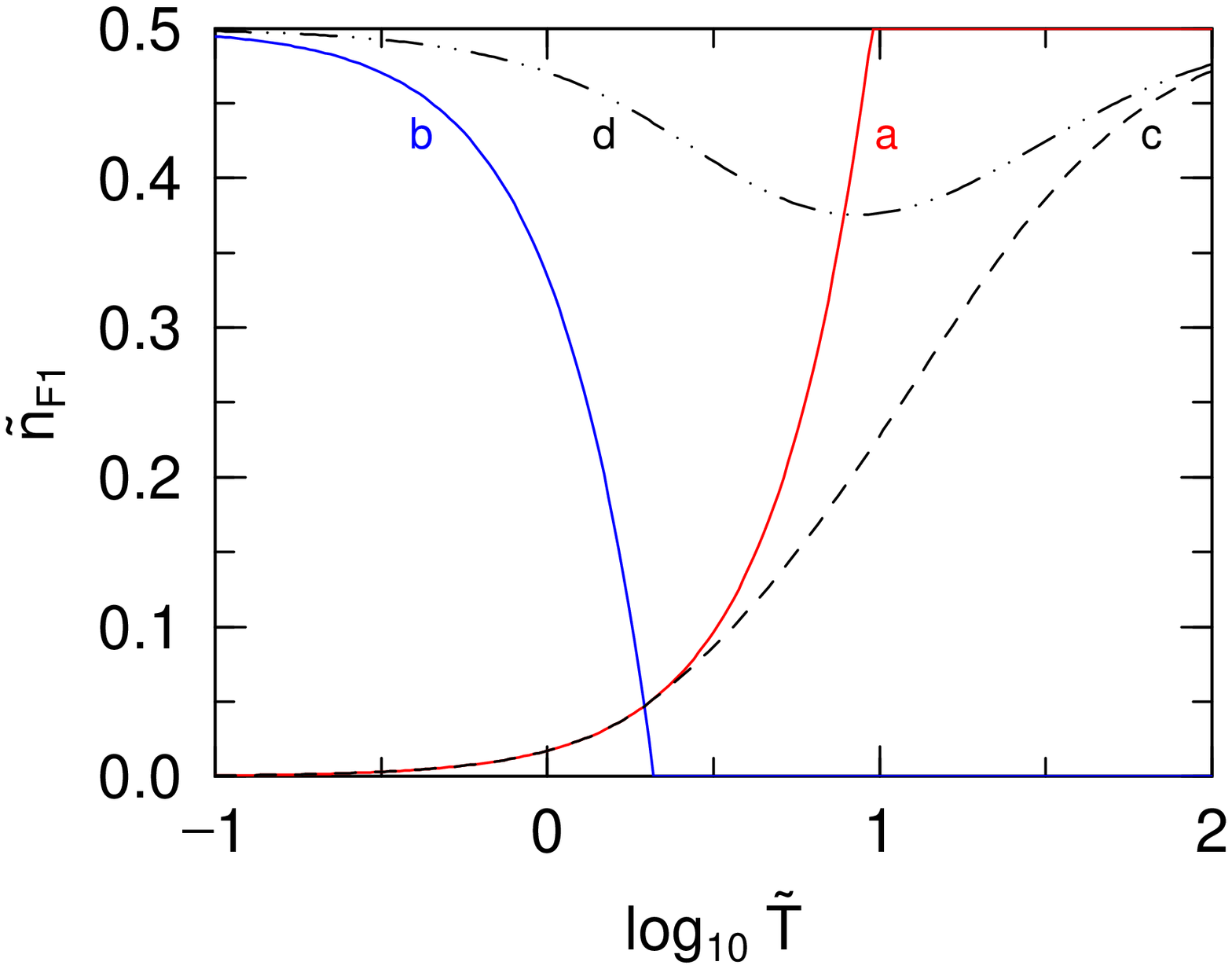} &
       \includegraphics[scale=0.3,angle=-90]{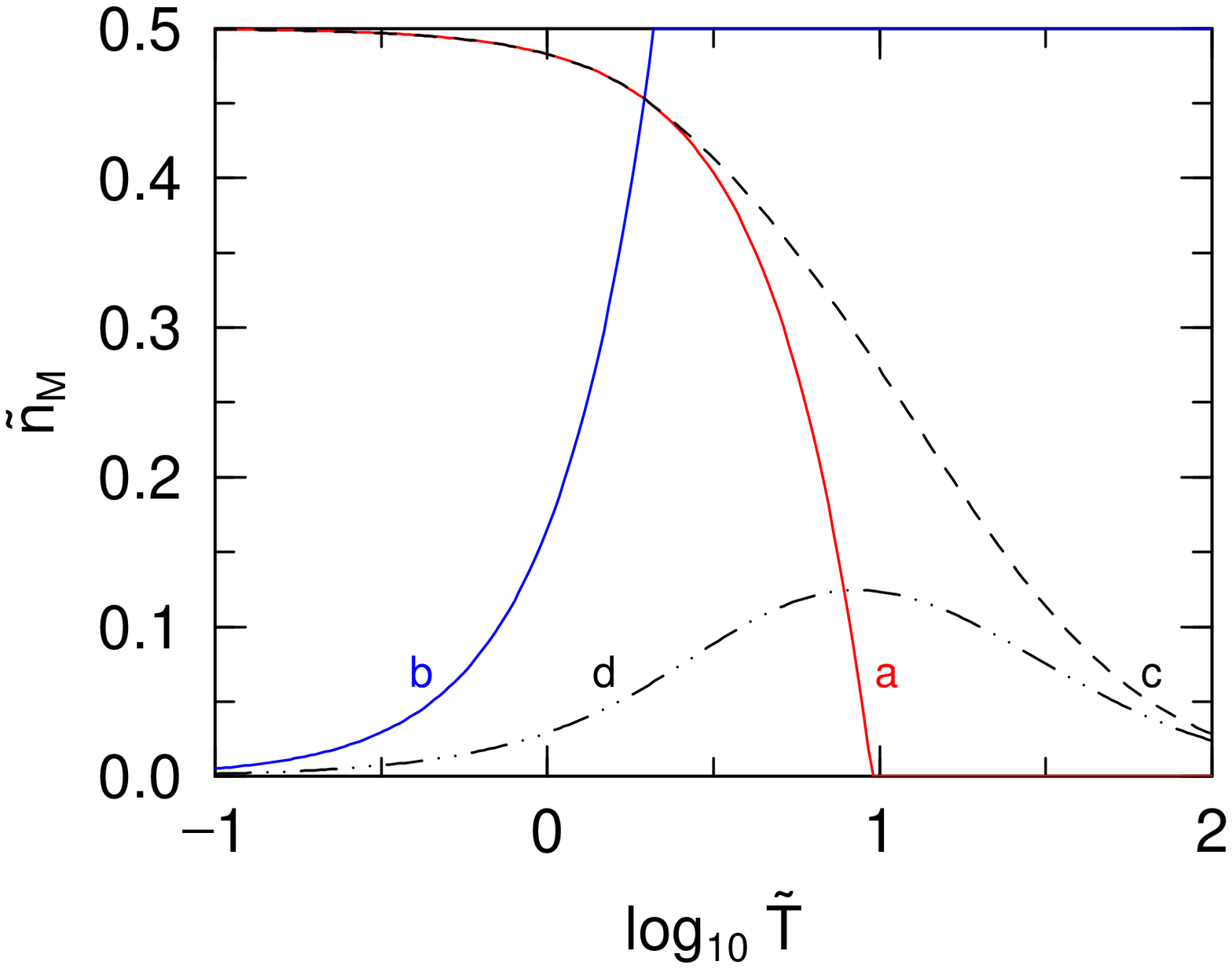}
    \end{tabular}
  \end{center}
\caption{(Color online) 
Temperature dependences of the number densities 
$\nT_{F1}$ of the atom $F1$ [(a), left] 
and $\nT_M$ of the molecule $M$ [(b), right] 
in the FF mixture 
with the same atomic masses and the same total atomic densities.
The lines $c$ (dashed line) and $d$ (double-dot-dashed line) are 
for mixtures with 
$\Delta{\ET_M} =0$ and $19.1$
(resonance state), respectively. 
The $\muT_F=0$ lines and the BEC $T_C$ lines are drawn 
for $F1$ and $F_2$ (line $a$) 
and $M$ (line $b$).
}
\label{fig5}
\end{figure}

The equilibrium states at $T \neq 0$ 
can be expressed in the same way as in the BF mixture. 
In Fig.~\ref{fig5}, we show the temperature dependences of 
$\nT_{F1}$ and $\nT_M$ in the FF mixture 
with $\nT_{F1,t} =\nT_{F2,t}=1/2$
(the states on the line $\nT_{F1,t}=1/2$ in Fig.~\ref{fig4})
and $\Delta\ET_M =0$ and $19.1$. 
In the FF mixture, 
the $T=0$ phases are classified with the value of $\Delta\ET_M$:
(i) $\Delta\ET_M <0$ (molecule phase), 
(ii) $0< \Delta\ET_M < \Delta_3$ 
(mixed phase), and 
(iii) $\Delta_3 <\Delta\ET_M$ 
(dissociated phase), 
where the border points with $\Delta\ET_M =0$ and
$\Delta_3 \equiv 2(3\pi^2)^{2/3} \approx 19.1$ 
correspond to point $A$ and the crossing point 
between line $BC$ and $\nT_{F1,t}=1/2$
in Fig.~\ref{fig4}.
The two lines $c$ and $d$ in Fig.~\ref{fig5} 
are for the border between two phases.

We should comment on the relation of the present approach
to the FF mixture with the BEC-BCS crossover theory\cite{Ohashi,Falco,Chen}. 
In the crossover theory, 
two kinds of bare fermions become 
dressed quasifermions, 
and quasimolecule states 
(or Cooper-pair states)
appear as physical degrees of freedoms.  
The strength change of the attraction between bare fermions 
gives the crossover between the BCS states 
(weak interaction) and the molecular BEC states 
(strong interaction). 
In the present quasiequilibrium approach, 
the fermions and the molecule should be considered 
as quasi-particles; 
the effects of bare-particle interactions 
are included as the existence of the molecule 
and its binding energy $\Delta\ET_M$. 
Really, the present approach 
is proved to give an approximated result 
on the strong interaction (BEC) side 
($\Delta\ET_M \lesssim 0$) \cite{Ohashi}.
In the weak interaction (BCS) side, 
the BEC $\TT_C$ in the present approach
vanishes at $\Delta\ET_M = \Delta_3$ 
as shown in Fig.~\ref{fig4} 
and disagrees with the BCS $T_C$. 
This is because the molecule states 
are treated as structureless bosons
and no statistical correlations are introduced. 
It can be said that the present approach 
should give a good approximation in the strong interaction side 
and in the high-$T$ region. 

\subsection{Molecular formations in the BB mixture}

Before discussing the results, 
we should give some explanations of the singular property 
of the boson chemical potential at the double limit of $n_B =0$ and $T=0$ 
in the no BEC region.

Around the point of $(n_B,T) \sim (0,0)$, 
$\nu =-\mu_B/k_B T$ becomes large, 
so that the Bose function $B_{3/2}$ in the density formula (\ref{EqB10}) 
can be approximated by an asymptotic formula
[Eq.~(\ref{EqApA9}) in Appendix A]: 
then the chemical potential becomes 
\begin{equation}
     \mu_B = k_B T \ln{(\lambda_T^3 n_B)} 
\label{EqC10}
\end{equation}
with $\lambda_T \propto T^{-1/2}$ 
defined in (\ref{EqB12}). 
Eq.~(\ref{EqC10}) shows that $\mu_B$ 
is singular at $(n_B,T) \sim (0,0)$ 
and can take any negative value $\mu_0$ 
if we take the limit through the pass 
$n_B \sim (\lambda_{T,B})^{-3} e^{\mu_0/k_B T}$.

\begin{figure}[ht]
  \begin{center}
       \includegraphics[scale=0.3,angle=-90]{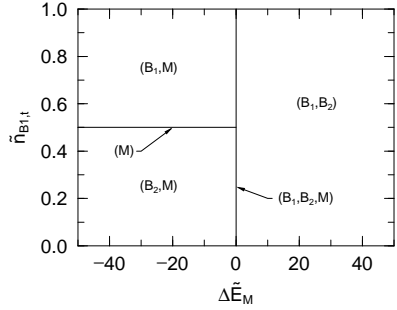}
  \end{center}
\caption{
The $T=0$ phase diagram of BB mixtures 
in the $\Delta{\ET_M}$-$\nT_{B1,t}$ plane.
The bracketed letters show 
what kinds of particles exist in equilibrium at $T=0$.
}
\label{fig7}
\end{figure}

Now let us go ahead to the atom-molecule equilibrium 
in the BB mixture 
through the process $B1 +B2 \leftrightarrow M =(B1 B2)$; 
then, the equilibrium condition (\ref{EqB17}) becomes
\begin{equation}
     \muT_{B1} +\muT_{B2} -\muT_M =\Delta\ET_M.
\label{EqC12}
\end{equation}
If all bosons condense into the BEC, 
the chemical potentials vanish, $\mu_{B1,B2,M} =0$, at $T=0$; 
it gives a contradiction in (\ref{EqC12}) 
except for the case of $\Delta\ET_M =0$ 
(no triple BEC theorem)\cite{Nawa}. 
The negativeness of the chemical potentials 
determines the solutions of (\ref{EqC12}) at $T=0$:
(i) $\Delta\ET_M <0$ (molecular phase): 
$\muT_M =0$ (the BEC of $M$) 
and $\muT_{\alpha} =\Delta\ET_M$ with $\nT_{\alpha} =0$ 
($\alpha = B1$ or $B2$). 
(ii) $\Delta\ET_M >0$ (dissociated phase): 
$\muT_{B1} =\muT_{B2} =0$ 
(the BECs of $B1$ and $B2$) and $\muT_M =-\Delta\ET_M$ 
with $\nT_M =0$. 
The results are summarized in the $T=0$ phase diagram 
in Fig.~\ref{fig7}. 
We should find that no triple BEC states exist 
except for the states on the boundary line of $\Delta\ET_M =0$.
Accordingly no mixed phase exists in the BB mixture.

\begin{figure}[ht]
  \begin{center}
    \begin{tabular}{cc}
       \includegraphics[scale=0.3,angle=-90]{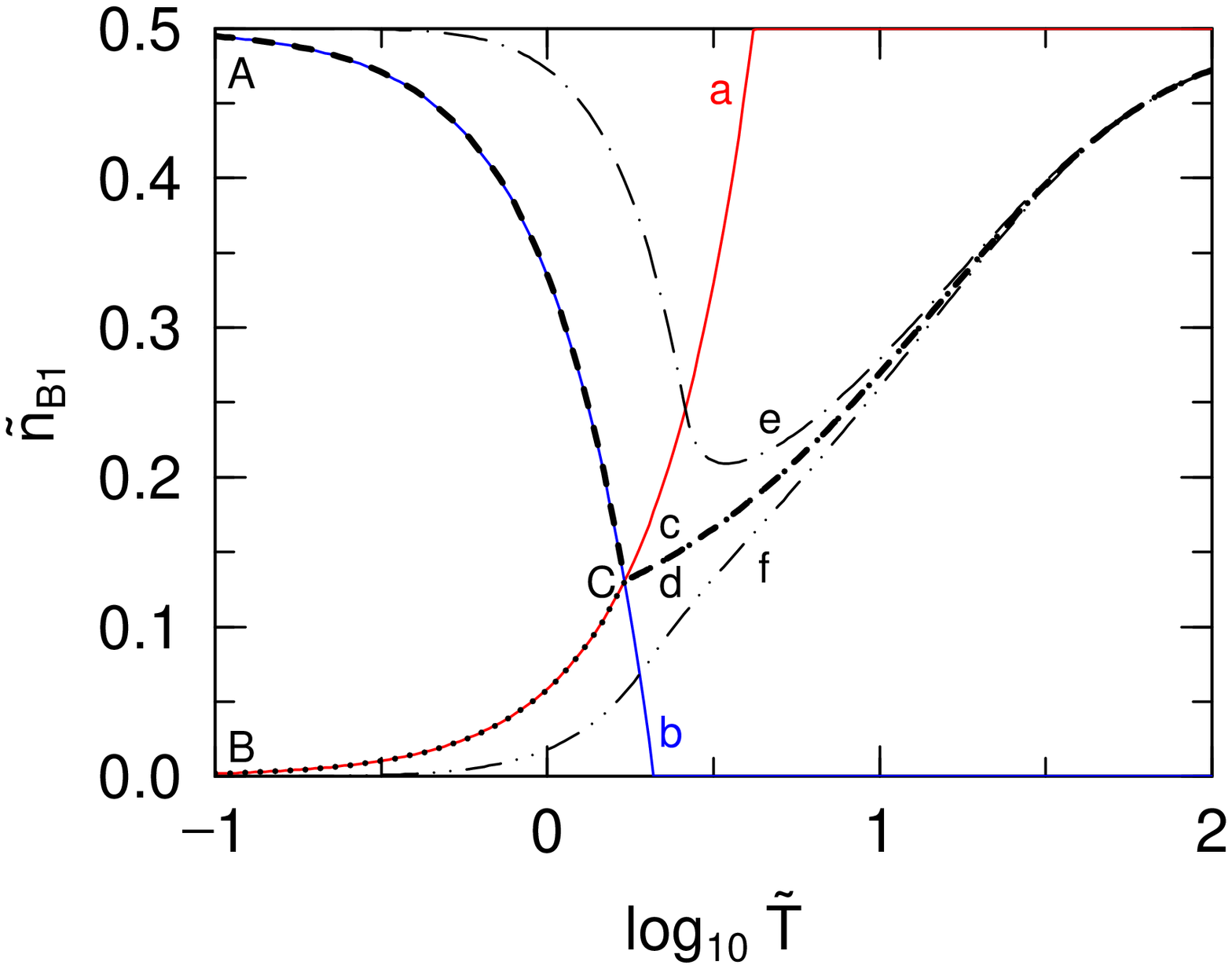} &
       \includegraphics[scale=0.3,angle=-90]{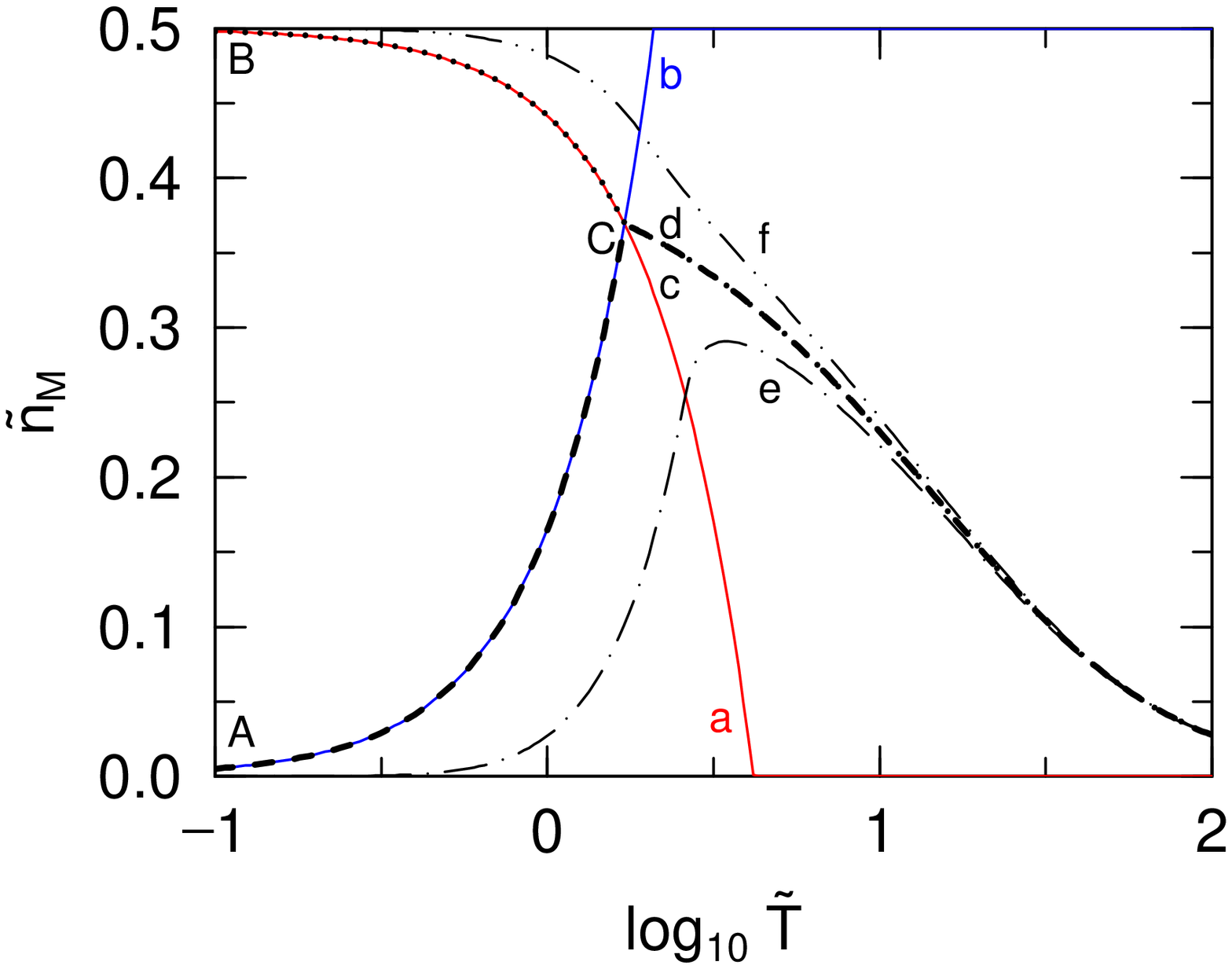}
    \end{tabular}
  \end{center}
\caption{(Color online) 
Temperature dependences 
of the number densities 
$\nT_{B1}$ of the atom $B1$ [(a), left] and 
$\nT_M$ of the molecule $M$ [(b), right] 
in the BB mixture 
with the same atomic masses and the same total atomic densities.
The lines $c$ (short-dashed line), $d$ (dotted line), $e$ (dot-dashed line) and $f$ (double-dot-dashed line) are for mixtures with 
$\Delta{\ET_M} =+0$, $-0$, $+1$, and $-1$, respectively, 
where 
$\Delta\ET_M =\pm 0 \equiv \lim_{\epsilon \to +0} \pm \epsilon$.
The BEC $T_C$ lines of $B1$ and $B2$ (line $a$) 
and $M$ (line $b$) are also drawn. 
}
\label{fig8}
\end{figure}

Fig.~\ref{fig8} shows the temperature dependences of 
$\nT_{B1}$ and $\nT_M$ in the BB mixture 
with $\mT_{B1} =\mT_{B2} =1/2$ 
and $\nT_{B1,t} =\nT_{B2,t}=1/2$
(the states on the line $\nT_{B1,t}=1/2$ in Fig.~\ref{fig7})
for $\Delta\ET_M = \pm 0, \pm 1$. 
The above-mentioned exclusive behaviors at $T=0$ for BEC 
have influences also at $T \neq 0$ 
as shown in Fig.~\ref{fig8}: 
in $\Delta\ET_M <0$, 
only the molecules become the BEC (lines $e$); 
otherwise, 
only atomic BECs can occur (lines $f$).
The mixture with $\Delta\ET_M =0$ has a singularity 
at point $C$; 
if we take the different kinds of limit 
$\Delta\ET_M = \pm 0$, 
BEC of atoms or molecules occurs, respectively.  

The exclusiveness 
between atom and molecule BECs in BB mixtures 
has been observed in a Cs experiment by the Innsbruck group\cite{Jochim2};
a sudden interchange occurs between the atom and molecule BECs 
when the resonance becomes the bound molecule 
using the Feshbach resonance method. 
This observation shows that the exclusiveness 
is incomplete and some mixture of atom and molecule BECs 
exists on the bound molecule side; 
it might be explained by 
an interparticle interaction effect
(see Sec.~V)
or nonequilibrium effect. 

%
\section{Law of Quantum Mass Action}
%

In mixtures of classical Maxwell-Boltzmann gases, 
the molecular formation or dissociation processes 
through (\ref{EqB1}) satisfy the ``law of mass action'':
\begin{equation}
     \frac{n_1 n_2}{n_M} =K(T),  \label{EqD1}
\end{equation}
where $n_{1,2,M}$ are the number densities of the particle 
$A_{1,2,M}$ 
and $K(T)$ is an equilibrium constant, 
which depends on $T$, but not on $n_{1,t}$ and $n_{2,t}$. 

In the present calculations 
of ultracold molecular formation or dissociation processes,
quantum-statistical effects play an important role, 
so that they give the density dependence of $K(T)$ in (\ref{EqD1});
we call it the ``law of quantum mass action''.

In order to make a comparison with the quantum cases, 
we derive an exact form of (\ref{EqD1}) 
in the case of ideal Maxwell-Boltzmann gases, 
where the density of the particle $A_\alpha$ is given by
\begin{equation}
     n^{(MB)}_\alpha = \lambda_{T,\alpha}^{-3}~ e^{\mu_\alpha/k_B T}, 
\label{EqD2}
\end{equation}
with the thermal de~Broglie wave length defined in (\ref{EqB12}).
Substituting Eq.~(\ref{EqD2}) into the equilibrium condition 
(\ref{EqB2}), 
we obtain the classical law of mass action~\cite{Landau,Laidler}:
\begin{equation}
     \frac{n^{(MB)}_1 n^{(MB)}_2}{n^{(MB)}_M} 
     =\left( \frac{\lambda_{T,M}}{\lambda_{T,1} \lambda_{T,2}} \right)^{3} 
     e^{\Delta{G}/k_B T} \equiv K(T),
\label{EqD3}  
\end{equation}
where $\Delta{G}$ 
is called the standard chemical affinity, 
and it becomes $\Delta{G} = \Delta{E_M}$ 
(the binding energy of the molecule $A_M$).

To show the deviations from the law of quantum action 
by quantum effects,
we evaluate the ratio of $n_1 n_2/n_M$ 
calculated in the previous section (including the quantum effects) 
to $n^{MB}_1 n^{MB}_2/n^{MB}_M =K(T)$ given by (\ref{EqD3}):
\begin{equation}
     R \equiv \frac{n_1 n_2}{n_M} \frac{n^{(MB)}_M}{n^{(MB)}_1 n^{(MB)}_2} 
       =\frac{n_1 n_2}{n_M} \frac{1}{K(T)}.
\label{EqD5}
\end{equation}
The ratio $R$ generally depends on $\TT$ and $\nT_{1,t}$
because of the constraint (\ref{EqB15}).

\begin{figure}[ht]
  \begin{center}
    \begin{tabular}{cc}
       \includegraphics[scale=0.3,angle=-90]{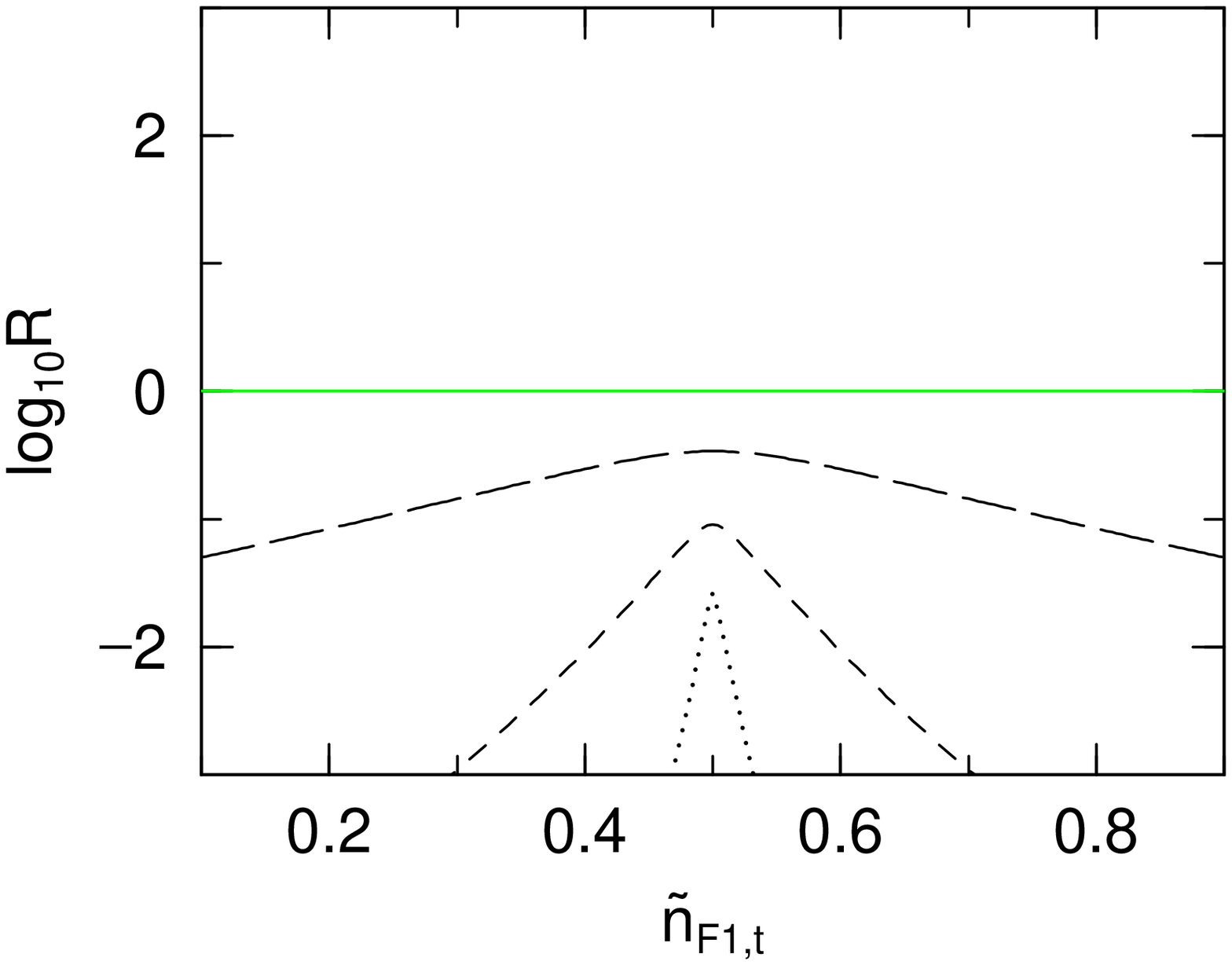} &
       \includegraphics[scale=0.3,angle=-90]{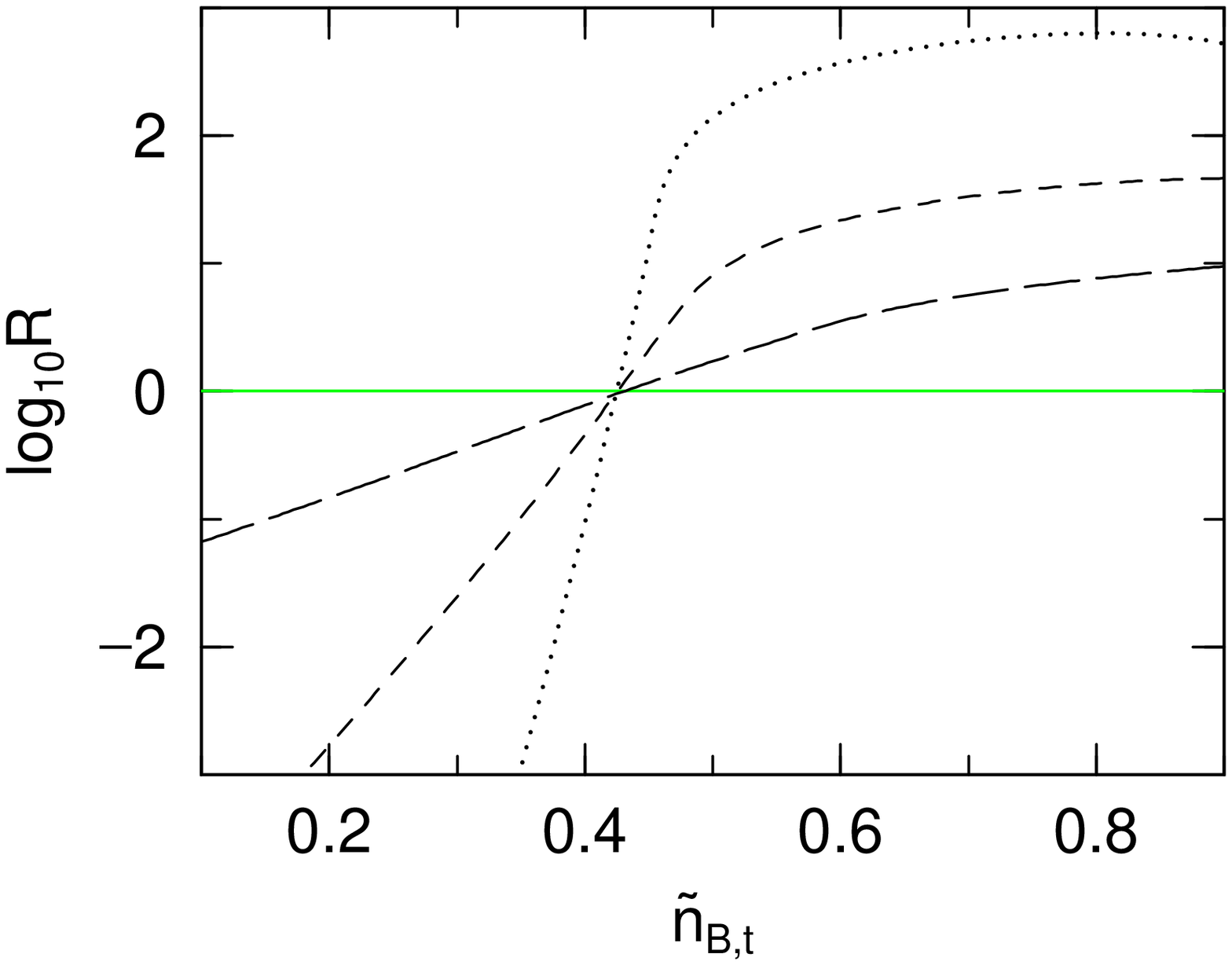} \\
       \includegraphics[scale=0.3,angle=-90]{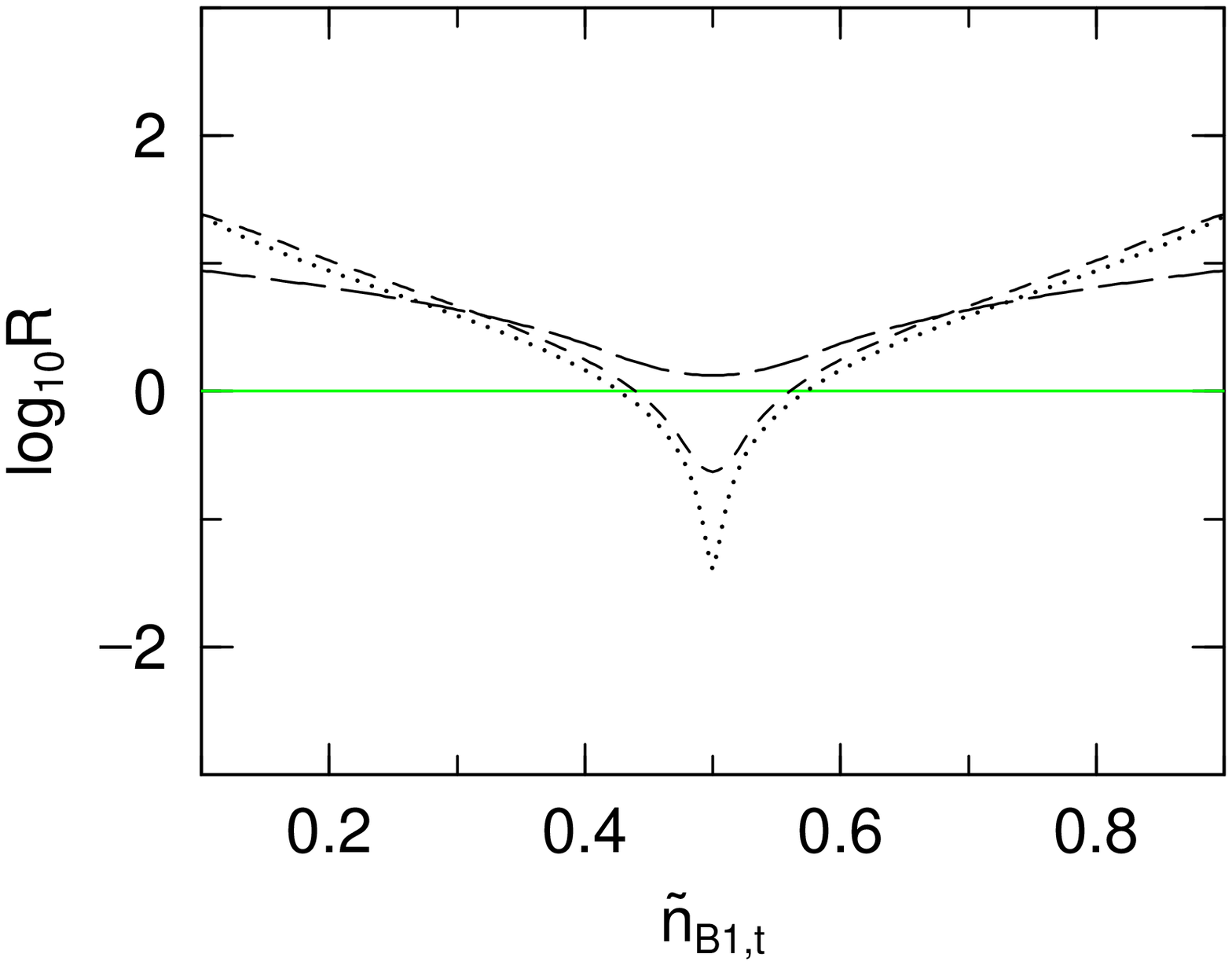} &
       \includegraphics[scale=0.3,angle=-90]{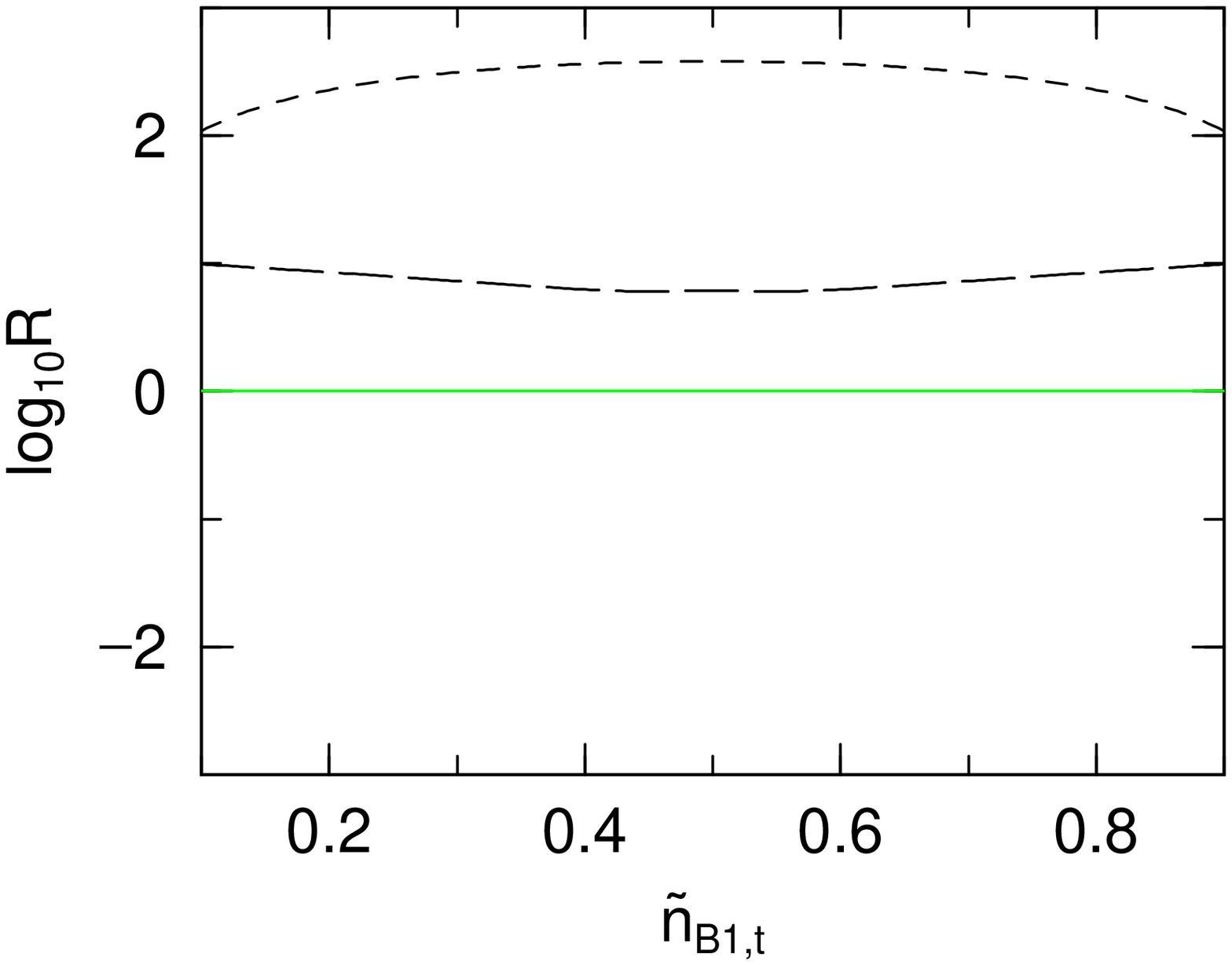} \\
    \end{tabular}
  \end{center}
\caption{(Color online) 
The ratio $R$ defined by (\ref{EqD5}) 
for the FF [(a), top left], BF [(b), top right], and 
BB [(c) and (d), bottom left and right] mixtures with the same atom masses.
The mass defects are $\Delta{\mT_M}=-1$ 
for the FF and BF mixtures 
and $\Delta{\mT_M}= -1$ (c), $+1$ (d) for the BB mixtures.
The lines are for temperatures: 
$\log_{10}\tilde{T} =\infty$ (Maxwell-Boltzmann limit, solid line),  
$0.4$ (long-dashed line), $0$ (short-dashed line), and $-0.4$ (dotted line), respectively.
}
\label{fig10}
\end{figure}

In Fig.~\ref{fig10}, 
we show the ${\tilde n}_{1,t}$ dependences of 
$\log{R}$ for several temperatures ($\log\TT =-0.4, 0, 0.4, \infty$) 
in the case of FF, BF, and BB mixtures 
with $\mT_1 =\mT_2 =1/2$.  
The mass defects are $\Delta{\mT_M}=-1$ 
for the FF and BF mixtures 
and $\Delta{\mT_M}=+1$ [Fig.~\ref{fig10}(c)] 
and $-1$ [Fig.~(\ref{fig10}(d)] for the BB mixtures. 
In the BB mixture, the behaviors of $R$ are found to be very different 
for molecules and resonances.

To understand the deviations from the law of mass action,
we expand the ratio $R$ in the high-$T$ region analytically:
\begin{eqnarray}
     R_{BB} &\sim& 1 + \frac{1}{2^{3/2}} \left( - \lambda_{T,1}^{3} n_{1} - \lambda_{T,2}^{3} n_{2} + \lambda_{T,M}^{3} n_{M} \right),  
\label{EqD6}\\
     R_{FF} &\sim& 1 + \frac{1}{2^{3/2}} \left( + \lambda_{T,1}^{3} n_{1} + \lambda_{T,2}^{3} n_{2} + \lambda_{T,M}^{3} n_{M} \right),  
\label{EqD7}\\
     R_{BF} &\sim& 1 + \frac{1}{2^{3/2}} \left( - \lambda_{T,1}^{3} n_{1} + \lambda_{T,2}^{3} n_{2} - \lambda_{T,M}^{3} n_{M} \right),  
\label{EqD8}
\end{eqnarray}
where $R_{BB,FF,BF}$ are for the BB,FF,BF mixtures.
The leading-order terms $R \sim 1$ in Eqs.~(\ref{EqD6})-(\ref{EqD8}) 
correspond to the Maxwell-Boltzmann limit (law of mass action). 
Completely different contributions are found 
in the next-order terms ($\propto T^{-3/2}$) 
for $R_{BB}$ and $R_{FF}$ 
in BB and FF mixtures, 
which are consistent with the results 
in Fig.~\ref{fig10}(a) and Figs.~\ref{fig10}(c) and \ref{fig10}(d) 
at high $T$.
The relatively small $T$ dependence of the ratio $R_{BF}$ 
for the BF mixture [Fig.~\ref{fig10}(b)] 
at high $T$ can be explained quantitatively 
from the cancellation 
of the boson and fermion contributions in the $T^{3/2}$ term
in (\ref{EqD8}).

%
%
\section{Effects of Interparticle Interactions on Equilibrium}
%
%
%
%
\subsection{Interaction effects in the mean-field approximation}

In the previous sections, 
we discussed atom-molecule equilibrium 
in noninteracting atom-gas mixtures. 
If interactions exist between atom-atom, atom-molecule, and molecule-molecule
(interacting) mixtures,
they should modify 
the equilibration 
and sufficiently strong interactions could change the phase structures of the mixtures. 

Theoretically, the effects of the interactions for the free energy 
can be divided into two kinds: 
mean-field and correlation ones. 
The mean-field effect can be evaluated, 
for example, using the Hartree-Fock approximation, 
and it can be represented as interaction terms 
(background energy)
in the one-particle energies of atoms or molecules 
in the mixtures. 
The correlation effects are defined as the contributions 
that cannot be introduced in the mean-field approximation. 
It should be noted that this division of interaction effects  
is a theoretical one and sometimes ambiguous. 
For example, in the BCS theory of superconductors, 
if we take the normal quasielectron theory 
of a normal Fermi-degenerate vacuum 
in the Hartree-Fock sense, 
then Cooper pairs and their condensations
are created by correlation effects; 
however, if we take the Bogoliubov quasi-particles 
as dynamical degrees of freedom, 
the BCS states can be understood as a kind of mean-field theory. 
In this paper, 
we discuss interaction effects for atom-molecule equilibrium 
in the Hartree-Fock-like mean-field approximation.  

For interparticle interactions, 
we take the ones 
coming through the two-body $s$-wave scattering processes, 
which give dominant contributions in ultracold atomic gases 
except the strongly interacting or spin-degenerate ones.
They can be introduced 
through effective interactions, 
for which we use the pseudopotentials\cite{Huang} 
\begin{equation}
     V_{i,j} =g_{i,j} 
              \sum_{\alpha,\beta} 
              \delta^3({\bf r}_\alpha -{\bf r}'_\beta), 
     \qquad (i,j =A1,A2,M)
\label{EqE1}
\end{equation}
between the $\alpha$th and $\beta$th particles 
of the $i$ and $j$ species, respectively.
The coupling constants $g_{i,j}$ are determined 
from the $s$-wave scattering lengths $a_{i,j}$ 
between $i$ and $j$ species:
\begin{equation}
     g_{i,j} =\frac{2\pi \hbar^2}{\mu_{i,j}} a_{i,j},
\label{EqE2}
\end{equation}
where $\mu_{i,j} \equiv m_i m_j/(m_i+m_j)$ 
is the reduced mass.

Let us consider the contributions 
of the interactions in mixtures 
with equilibrium $A1 +A2 \leftrightarrow M$ 
in the mean-field approximation.
 
The interaction effects are introduced 
into the free energy as the background energy: 
\begin{equation}
     F =E_0 +E^{int} -\sum_{i=A1,A2,M} \mu_i n_i,
\label{EqE3}
\end{equation}
where the $E_0$ is the kinetic energy, 
which exists also in noninteracting cases. 

The contribution of the potential $V_{i,j}$ in (\ref{EqE1}) 
to the background energy $E^{int}$ 
is evaluated by
$E^{int}_{i,j} =\langle V_{i,j} \rangle$.
In the mean-field approximation, 
it is expressed by the number densities $n_i$ and $n_j$.
In the interaction between the same kinds of particles,
it becomes
\begin{equation}
     E_{BB} =\frac{g_{BB}}{2} 
             \left\{ 2 n_B^2 
                    - (n_B^{(0)})^2 \right\}, \qquad 
     E_{FF} =0,
\label{EqE4}
\end{equation}
for the boson $B$ and the fermion $F$. 
The $n_B^{(0)}$ is 
the condensed density of $B$; 
when $T \sim 0$, we can use the approximation 
that $n_B^{(0)} \sim n_B$. 
The vanishing background energy for the fermion $F$ 
originates in forbidden $s$-wave scatterings 
by Pauli blocking effects at $T \sim 0$.
To avoid unnecessary complexity in the formulation, 
we redefine $g_{FF}=0$ for the fermions.
The background energies coming from the interactions 
between different kinds of particles become
\begin{equation}
     E_{i,j} =g_{i,j} n_i n_j \qquad
     (i \neq j).
\label{EqE5}
\end{equation}

From the above considerations, 
the total background energy 
in the equilibrium through $A_1+A_2 \leftrightarrow M$ 
can be approximated by
\begin{equation}
     E^{int} =\sum_{i=A1,A2,M} \frac{g_{i,i}}{2} (n_i)^2
             +g_{A1,A2} n_{A1} n_{A2}
             +g_{A1,M} n_{A1} n_M
             +g_{A2,M} n_{A2} n_M.
\label{EqE6}
\end{equation}

Differentiating the free energy $F$ in (\ref{EqE3}) 
with respect to the density $n_i$, 
we obtain the single-particle energy
\begin{equation}
      \epsilon_i =\epsilon_i^{(0)} +\sum_{j=A1,A2,M} g_{i,j} n_j -\mu_i,
\label{EqE7}
\end{equation}
where $\epsilon_i^{(0)}$ is the kinetic energy of particle $i$:
\begin{equation}
     \epsilon_i^{(0)} =\frac{\partial E_0}{\partial n_i} 
                 =\frac{({\bf p}_i)^2}{2m_i}.
\label{EqE8}
\end{equation}
With the effective chemical potential defined by
\begin{equation}
     \mu'_i =\mu_i -\sum_{j=A1,A2,M} g_{i,j} n_j,
\label{EqE9}
\end{equation}
Eq.~(\ref{EqE7}) becomes
\begin{equation}
     \epsilon_i =\epsilon_i^{(0)} -\mu'_i.
\label{EqE10}
\end{equation}
It should be noticed that 
the $\epsilon_i$ in (\ref{EqE10}) has the same form 
as that in the noninteracting case, 
so that the same Bose and Fermi statistic formulas 
(\ref{EqB19}) and (\ref{EqB20})
can be applied also in the interacting case
using the effective chemical potential $\mu'_i$ 
instead of $\mu'_i$:
\begin{eqnarray}
     \nT_\alpha &=& \left[ \frac{\mT_\alpha \TT}{2\pi} 
                           \right]^{3/2}
                           B_{3/2}(-\muT'_\alpha/\TT) 
     \qquad \hbox{(for boson $A_\alpha$)}, \label{EqE11} \\ 
     \nT_\alpha &=& \left[ \frac{\mT_\alpha \TT}{2\pi} 
                           \right]^{3/2}
                           F_{3/2}(-\muT'_\alpha/\TT) 
     \qquad \hbox{(for fermion $A_\alpha$)}. \label{EqE12} 
\end{eqnarray}
Using eq.~(\ref{EqE9}), 
the equilibrium condition (\ref{EqB17}) for the interacting mixture becomes
\begin{equation}
      \mu'_{A1} +\mu'_{A2} -\mu'_M =\Delta{E'_M},
\label{EqE13}
\end{equation}
where the effective binding energy $\Delta{E'_M}$ 
is given by the density of the $M$ molecule
\begin{equation}
     \Delta{E'_M} =\alpha n_M +\gamma,
\label{EqE14}
\end{equation}
where interaction effects are included 
in the two parameters $\alpha$ and $\gamma$:
\begin{eqnarray}
     \alpha &=&\sum_{i=A1,A2,M} g_{i,i} 
             +2 (g_{A1,A2} -g_{A1,M} -g_{A2,M}), 
\label{EqE15}\\
     \gamma &=&\Delta{E} +(g_{A1,M} -g_{A1,A1} -g_{A1,A2}) n_{A1,t}
                       +(g_{A2,M} -g_{A2,A2} -g_{A1,A2}) n_{A2,t}.
\label{EqE16}
\end{eqnarray}
In the derivation of eqs.~(\ref{EqE14})-(\ref{EqE16}), 
we have used the constraints (\ref{EqB15}).

As has been done in the case of noninteracting mixtures in Sec.~IIB, 
we introduce scaled variables for the coupling constants 
and the effective binding energies:
\begin{eqnarray}
     \gT_{i,j} =\frac{g_{i,j} n_t}{E_s},  
\label{EqE17}\\
     \Delta{\ET'_M} =\frac{\Delta{\ET'_M}}{E_s},
\label{EqE18}
\end{eqnarray}
where $n_t =n_{1,t} +n_{2,t}$ 
and $E_s =\hbar^2(n_t)^{2/3}/m_M$.
For the effective binding energy $\Delta{\ET'_M}$, 
Eq.~(\ref{EqE14}) becomes
\begin{equation}
     \Delta{\ET'_M} =\gaT \nT_M +\gamT,
\label{EqE19}
\end{equation}
where the scaled parameters $\gaT$ and $\gamT$ are defined by
\begin{eqnarray}
     \gaT &=&\sum_{i=A1,A2,M} \gT_{i,i} 
             +2 (\gT_{A1,A2} -\gT_{A1,M} -\gT_{A2,M}), 
\label{EqE20}\\
     \gamT &=&\Delta{\ET} 
             +(\gT_{A1,M} -\gT_{A1,A1} -\gT_{A1,A2}) \nT_{A1,t}
             +(\gT_{A2,M} -\gT_{A2,A2} -\gT_{A1,A2}) \nT_{A2,t}.
\label{EqE21}
\end{eqnarray}
Using the scaled variables, 
the atom-molecule equilibrium condition (\ref{EqE13}) 
becomes
\begin{equation}
     \muT_1 +\muT_2 -\muT_M 
          =\Delta{\ET'_M}
          \equiv \gaT \nT_M +\gamT,
\label{EqE22}
\end{equation}
where Eq.~(\ref{EqE19}) has been used.

The atom-molecule equilibrium states 
of the interacting mixtures can be obtained 
from eqs.~(\ref{EqE11})-(\ref{EqE13}) 
with the constraints (\ref{EqB15}). 
The interaction effects are included 
in the effective binding energy $\Delta{E'_M}$ in (\ref{EqE14}) 
through the two parameters $\alpha$ and $\gamma$, 
which are determined from the coupling constants $g_{i,j}$. 
Thus, we find that, in the mean-field approximation, 
one extra parameter is necessary in
the equilibrium theory of interacting mixtures 
in comparison with that of noninteracting ones.

\subsection{Phase structure changes by interaction effects}

Before we give the numerical results 
for the atom-molecule equilibrium states 
of the interacting mixtures, 
we discuss qualitatively how the interactions shift and change 
the phase structures (PSs).
We assume that all coupling constants are positive $g_{i,j}>0$ 
to avoid possible instabilities
from the spatial fluctuations of densities\cite{Miyakawa1}.

Different from noninteracting cases 
where the equilibrium condition (\ref{EqB17}) has a unique solution, 
the existence of the density-dependent term $\gaT \nT_M$ in (\ref{EqE22})
results in two or more different solutions of (\ref{EqE22}), 
which correspond to the different equilibrium states
(coexisting phases).
These solutions include locally stable and unstable states, 
so that we have to examine the behavior of Eq.~(\ref{EqE22}) 
and take out solutions 
corresponding to stable states. 

Close examination shows that 
two critical points $0 > \gaT_{c1} > \gaT_{c2}$
exist in the parameter $\gaT$
for the BF, FF, and BB mixtures, 
and the $T=0$ equilibrium structures 
can be classified into three regions with them as folloes, 
\begin{description}
\item[PS1] ($\gaT_{c1} < \gaT$), 
Eq.~(\ref{EqE22}), has unique solutions 
for each value of the parameters $\gaT$, $\gamT$, and $\nT_{1,2,t}$, 
and the $T=0$ phase diagrams become similar in structure 
with those for the noninteracting mixtures,  
including the dissociate, mixed, and molecule phases. 
(As shown later, the BB mixture is somewhat exceptional. )
\item[PS2] ($\gaT_{c2} < \gaT < \gaT_{c1}$), 
Eq.~(\ref{EqE22}), has two stable (one unstable) solutions 
in the mixed phase.
\item[PS3] ($\gaT < \gaT_{c2}$) 
has 
new coexisting phases appearing 
where both the mixed and molecular states 
become locally stable.
\end{description}
 
Thus, interaction effects 
generally give complex phase structures 
in atom-molecule equilibrium 
with the phases with two locally stable states. 
The occurrence of these phases depends 
on the parameters $\gaT$, $\gamT$, and $\nT_{1,2,t}$,  
and the transitions of the states caused by the change of 
these parameters might be first order. 
In the phases with two locally stable states, 
the absolutely stable equilibrium state 
should be determined 
from a free-energy comparison of these states.
However, the difference of the free energies 
of these states is generally small, 
so that effects that are not included 
in the present mean-field calculations 
(the correlation effect) may give comparative 
contributions,  
and, also, in real experiments, 
states which are not absolutely stable 
can occur through nonequilibrium and history effects. 
For these reasons, 
we should say that the stability of these phases 
is very subtle, 
and a study of the more detailed structures of these phases 
will not be done in this paper. 

From Eq.~(\ref{EqE20}), 
the negative contributions to $\gaT$ 
can be given by large values of $\gT_{A1,M}$ and $\gT_{A2,M}$; 
thus, atom-molecule interactions are found to be interesting 
and important 
in transitions to unstable phases.

\subsection{Phase structures of the interacting BF mixture}

The critical values $\gaT^{(BF)}_{c1}$ and $\gaT^{(BF)}_{c2}$ 
for the BF mixture are given by 
\begin{eqnarray}
     \gaT^{(BF)}_{c1} &=& -\frac{2^{4/3}}{3} 
                           \left( \frac{3\pi^2}{\sqrt{2}} \right)^{2/3}
                           \left[ 1 +\frac{1}{2^{1/4} \mT_F^{3/4}} \right],
\label{EqE23}\\
     \gaT^{(BF)}_{c2} &=& -2^{1/3} 
                           \left( \frac{3\pi^2}{\sqrt{2}} \right)^{2/3}
                           \left[ 1 +\frac{1}{\mT_F} \right].
\label{EqE24}
\end{eqnarray}

\begin{figure}[t]
  \begin{center}
       \includegraphics[scale=0.4,angle=-90]{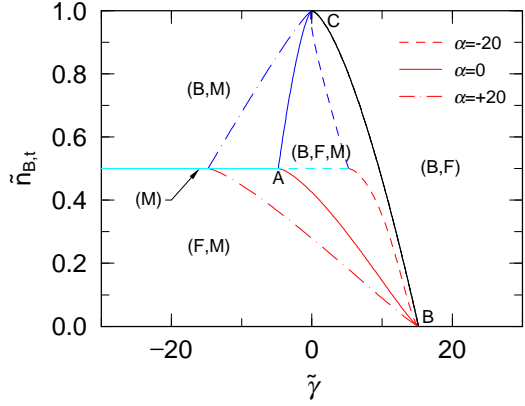}
  \end{center}
\caption{(Color online) 
The $T=0$ phase diagram of interacting BF mixtures 
with the same atom masses  
in the $\gamT$-$\nT_{B,t}$ plane for $\gaT =-20,0,20$.
The bracketed letters in the regions or lines show 
what kinds of particles exist in equilibrium at $T=0$.
}
\label{fig11}
\end{figure}

\begin{figure}[b]
  \begin{center}
       \includegraphics[scale=0.4,angle=-90]{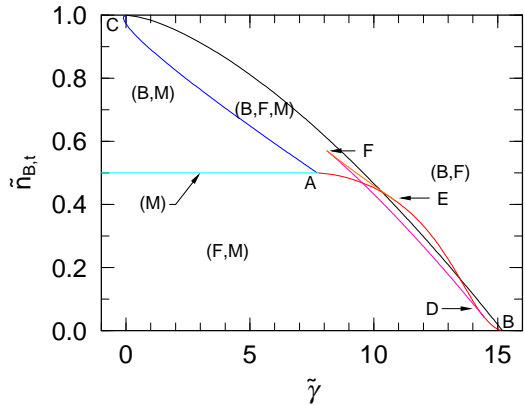}
  \end{center}
\caption{(Color online) 
The $T=0$ phase diagram of interacting BF mixtures 
with the same atom masses  
in the $\gamT$-$\nT_{B,t}$ plane for $\gaT =-25$.
The bracketed letters in the regions or lines show 
what kinds of particles exist in equilibrium at $T=0$.
}
\label{fig12}
\end{figure}

Fig.~\ref{fig11} shows the $T=0$ phase diagram 
for interacting BF mixtures 
with $\mT_B =\mT_F =1/2$ 
for $\gaT =-20,0,20$. 
Eqs.~(\ref{EqE23}) and (\ref{EqE24}) 
give $\gaT^{(BF)}_{c1} \sim -20.7$ and $\gaT^{(BF)}_{c2} \sim -28.7$
when $\mT_B =\mT_F =1/2$, 
so that the phase diagrams in Fig.~\ref{fig11} are found to show
the phase structure PS1;
they have qualitatively similar structures 
as the diagram for the noninteracting mixture (Fig.~\ref{fig1}).
Especially, the phase diagram for $\gaT=0$ (solid line in Fig.~\ref{fig11}) 
is completely the same as that in Fig.~\ref{fig1} 
because of the vanishing density-dependent term $\gaT \nT_M$ 
in (\ref{EqE22}). 
We also find that
the boundary between the dissociate and mixed phases, 
$CB$, in Fig.~\ref{fig11}, 
which is given by (\ref{EqApB16}) in Appendix B, 
is independent of the values of $\gaT$; 
this is because the $n_M =0$ condition on this boundary 
eliminates the $\gaT$ dependence in (\ref{EqE22}). 
The boundaries between the mixed and molecular phases 
depend on $\gaT$ 
as shown in (\ref{EqApB15}) and (\ref{EqApB17}) 
in Appendix B. 
The $\gaT$-dependence of the mixed phase 
can be understand from the position of the end point $A$ 
in Fig.~\ref{fig11}:  
\begin{equation}
     \gamT =-\left( \frac{3\pi^2}{2\sqrt{2}} \right)^{2/3}
             -\frac{\gaT}{2}
            \sim -4.78 -\frac{\gaT}{2},
\label{EqE25}
\end{equation}
which is given by (\ref{EqApB19}) in Appendix B.
Eq.~(\ref{EqE25}) shows that
the area of the mixed phase increases when $\gaT>0$ 
and decreases in the case of $\gaT<0$.
This behavior can be explained from (\ref{EqE22}); 
in the case of $\gaT>0$, 
the molecule density $\nT_M$ 
has the effect of increasing $\Delta{\ET'_M}$, 
which makes molecular formation difficult, 
and the area of the mixed phase becomes large
(the $\gaT <0$ case has the contrary effect).

\begin{figure}[bc]
  \begin{center}
    \begin{tabular}{cc}
       \includegraphics[scale=0.4,angle=-90]{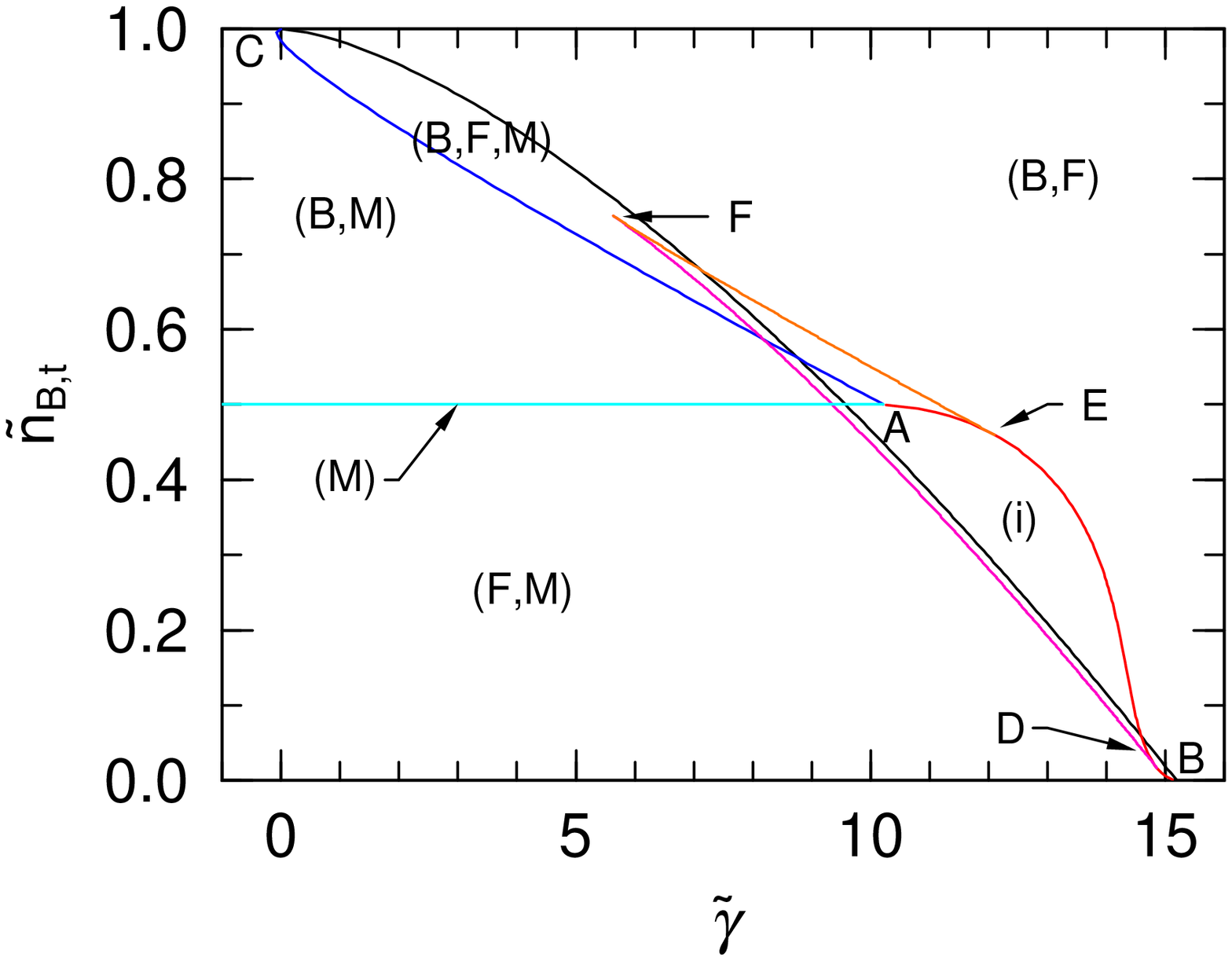} &
       \includegraphics[scale=0.4,angle=-90]{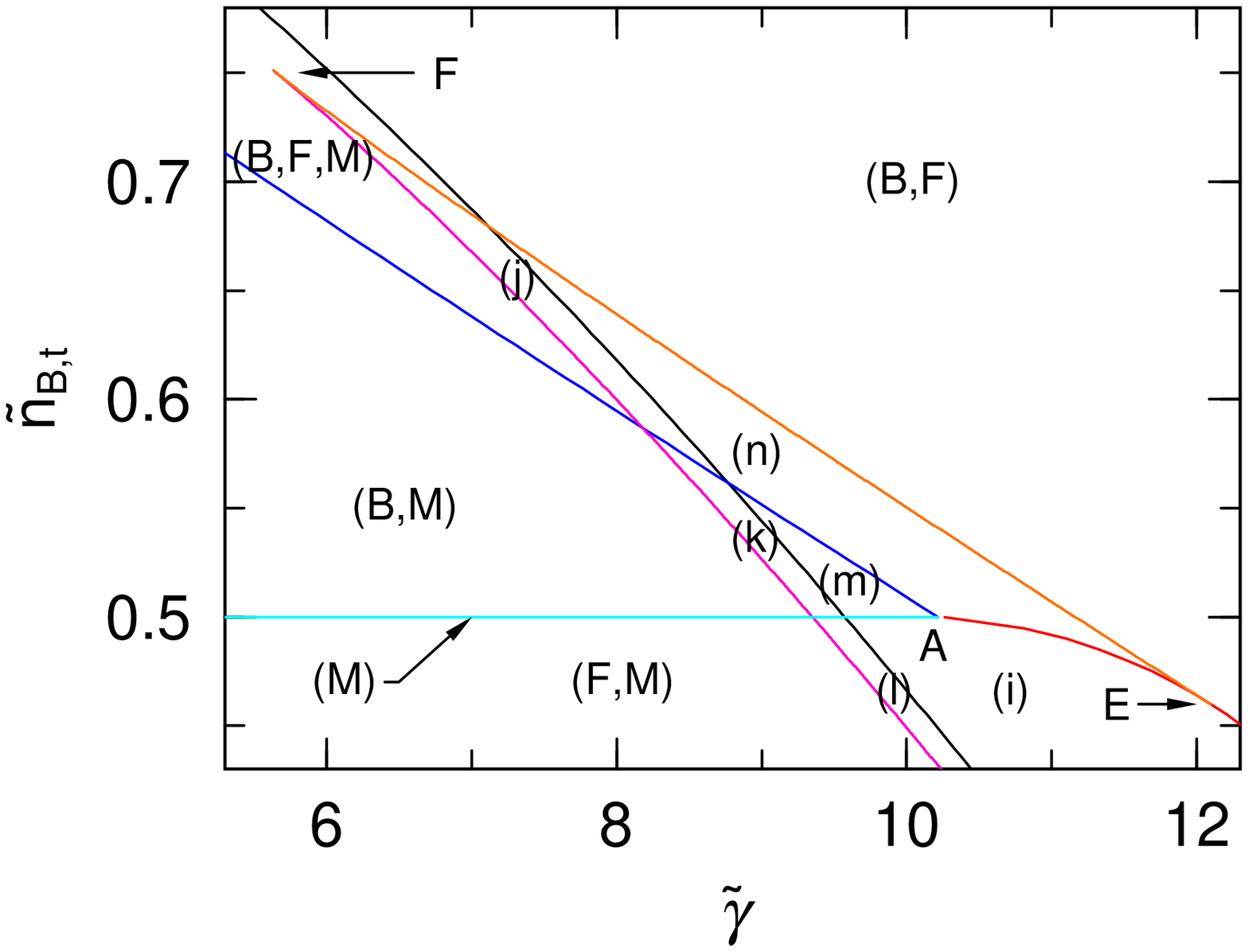} \\
       \includegraphics[scale=0.4,angle=-90]{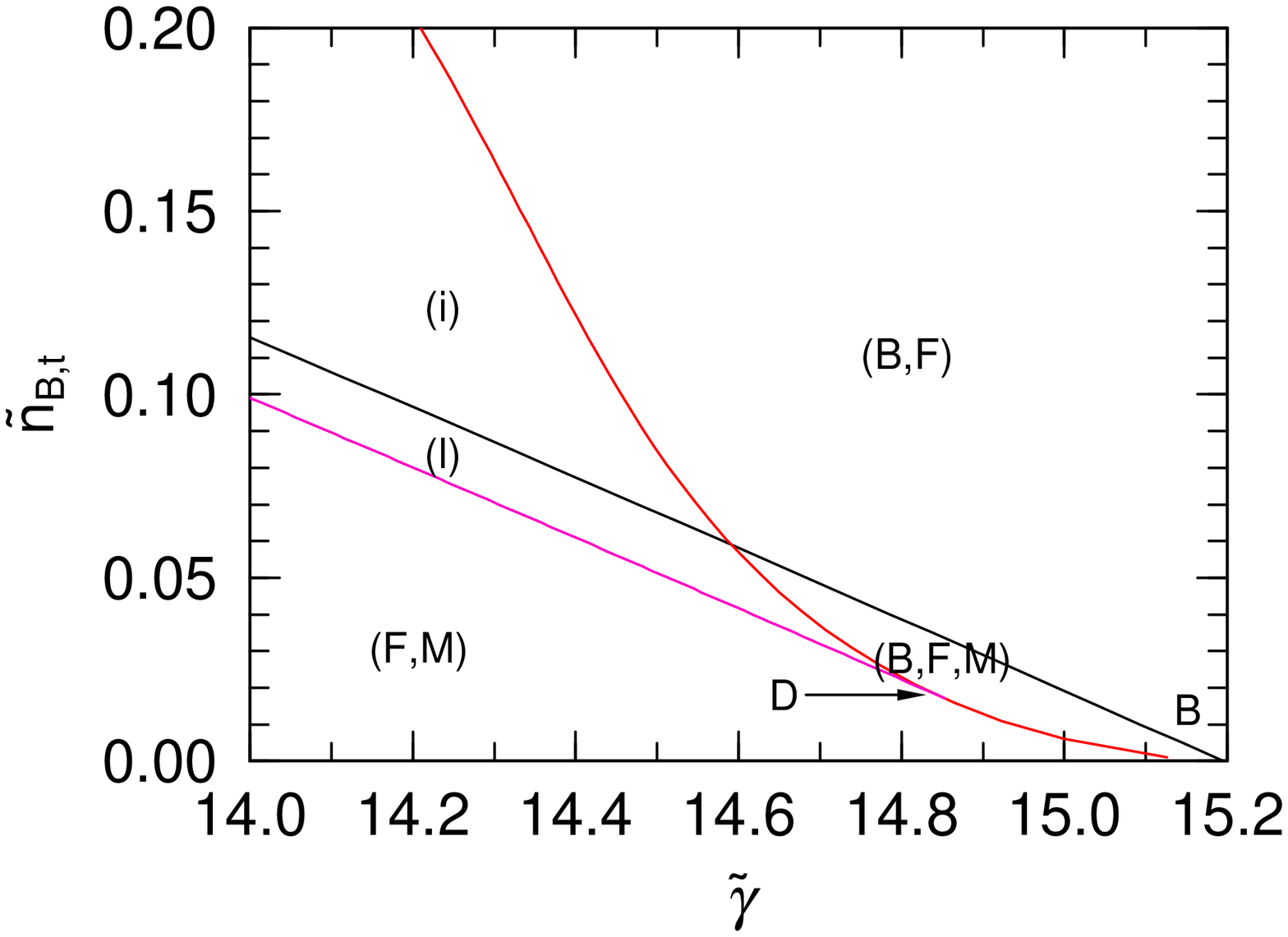} &
       \includegraphics[scale=0.4,angle=-90]{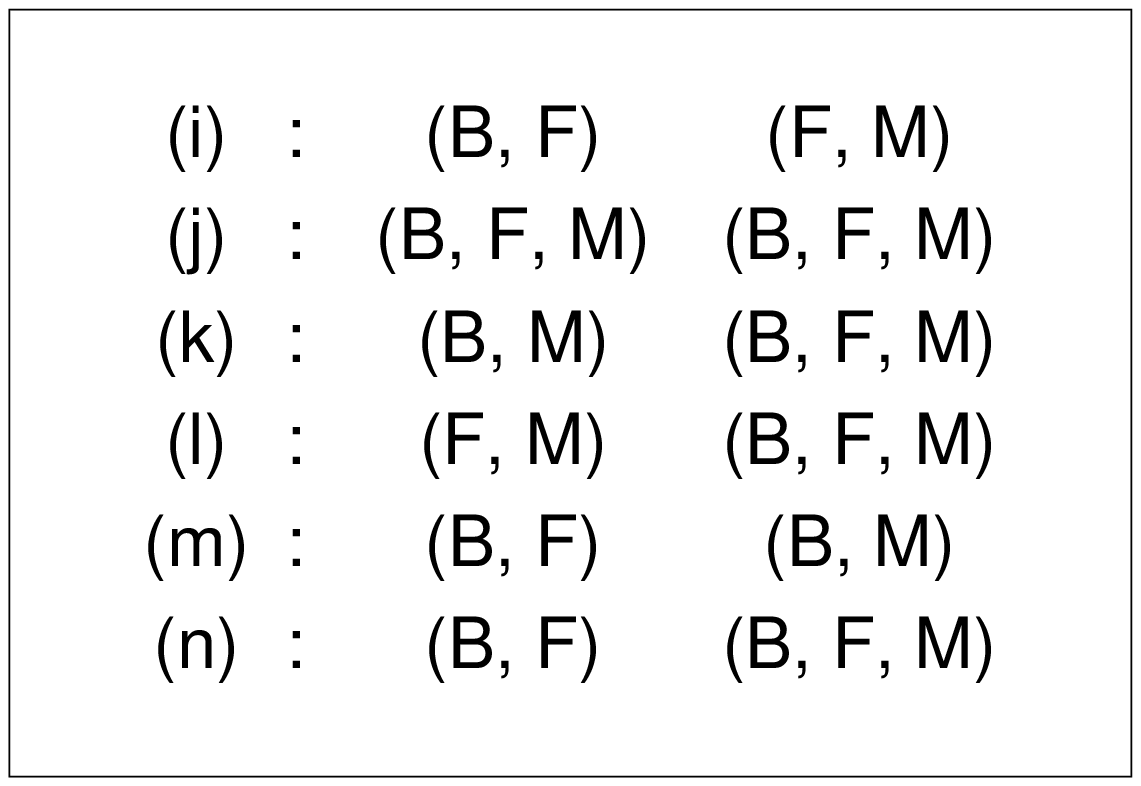}
    \end{tabular}
  \end{center}
\caption{(Color online) 
The $T=0$ phase diagram of interacting BF mixtures 
with the same atom masses 
in the $\gamT$-$\nT_{B,t}$ plane for $\gaT =-30$ [(a), top left]. 
(b) and (c) (top right and bottom left) show the detailed structures 
of (a).
(d) (bottom left) is the legend that explains the phases $(i)$-$(n)$ 
in (a)-(c).}
\label{fig13}
\end{figure}

\begin{figure}[t]
  \begin{center}
       \includegraphics[scale=0.4,angle=-90]{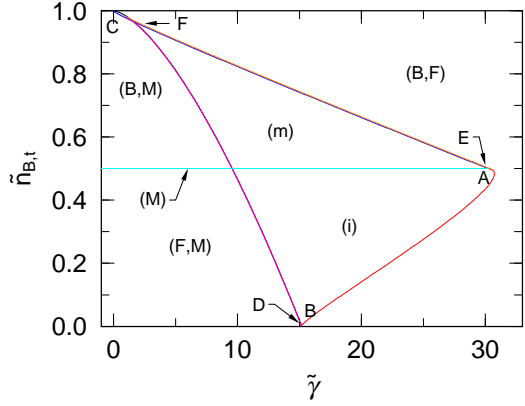}
  \end{center}
\caption{(Color online) 
The $T=0$ phase diagram of interacting BF mixtures 
with the same atom masses
in the $\gamT$-$\nT_{B,t}$ plane for $\gaT =-70$, 
where the phase $(i)$ and $(m)$ are coexisting phases 
of $(B, F)$, $(F, M)$ and $(B, F)$, $(B, M)$ 
as explained in the legend in Fig.~\ref{fig13}(d).}
\label{fig14}
\end{figure}

The phase diagrams with the phase structures PS2 and PS3 
are given in Fig.~\ref{fig12} ($\gaT =-25$) 
and Fig.~\ref{fig13} ($\gaT=-30$). 
In Fig.~\ref{fig12}, the occurrence of two locally stable states 
gives the additional two phases 
with the boundaries $(AB+DF)$ and $(DF+AB+EF)$. 
The areas of these new phases in the PS2 structure are small; 
they are nothing more than substructures in the mixed phase. 
In the PS3 structure (Fig.~\ref{fig13}), 
the left-shifted phase boundary $AB$ crosses 
the boundary $BC$, 
and new kinds of phase can be produced between $AB$ and $BC$: 
region (b) in Fig.~\ref{fig13}, for example, 
where dissociated and mixed states coexist.
Fig.~\ref{fig13} also has a very complex substructure 
around the boundary $BC$.
For extremely large values of $\gaT$ (Fig.~\ref{fig14}, $\gaT=-70$), 
the regions with substructures shrink to the small areas 
around the vertexes $B$ and $C$ in Fig.~\ref{fig14}, 
and the whole phase structure becomes simple again; 
the phases where the dissociated and molecular states coexist
appear in the central region, 
where the dissociated states include no molecule and 
the molecular states have as many molecules as possible. 

%
%
\subsection{Phase structures of interacting BB and FF mixtures}

In this subsection, we briefly sketch the structures 
of the $T=0$ phase diagrams 
of interacting BB and FF mixtures.

\begin{figure}[t]
  \begin{center}
    \begin{tabular}{cc}
       \includegraphics[scale=0.4,angle=-90]{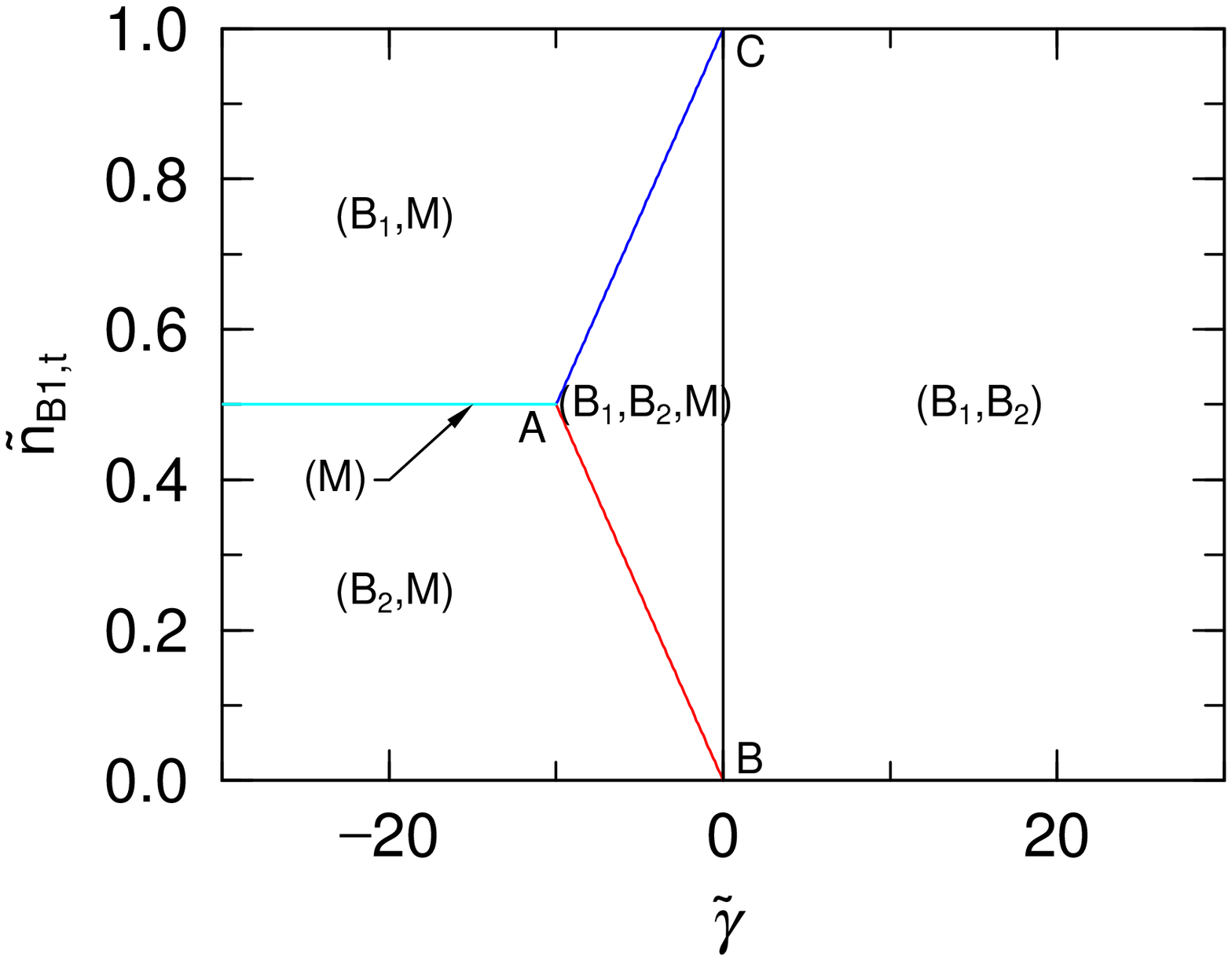} &
       \includegraphics[scale=0.4,angle=-90]{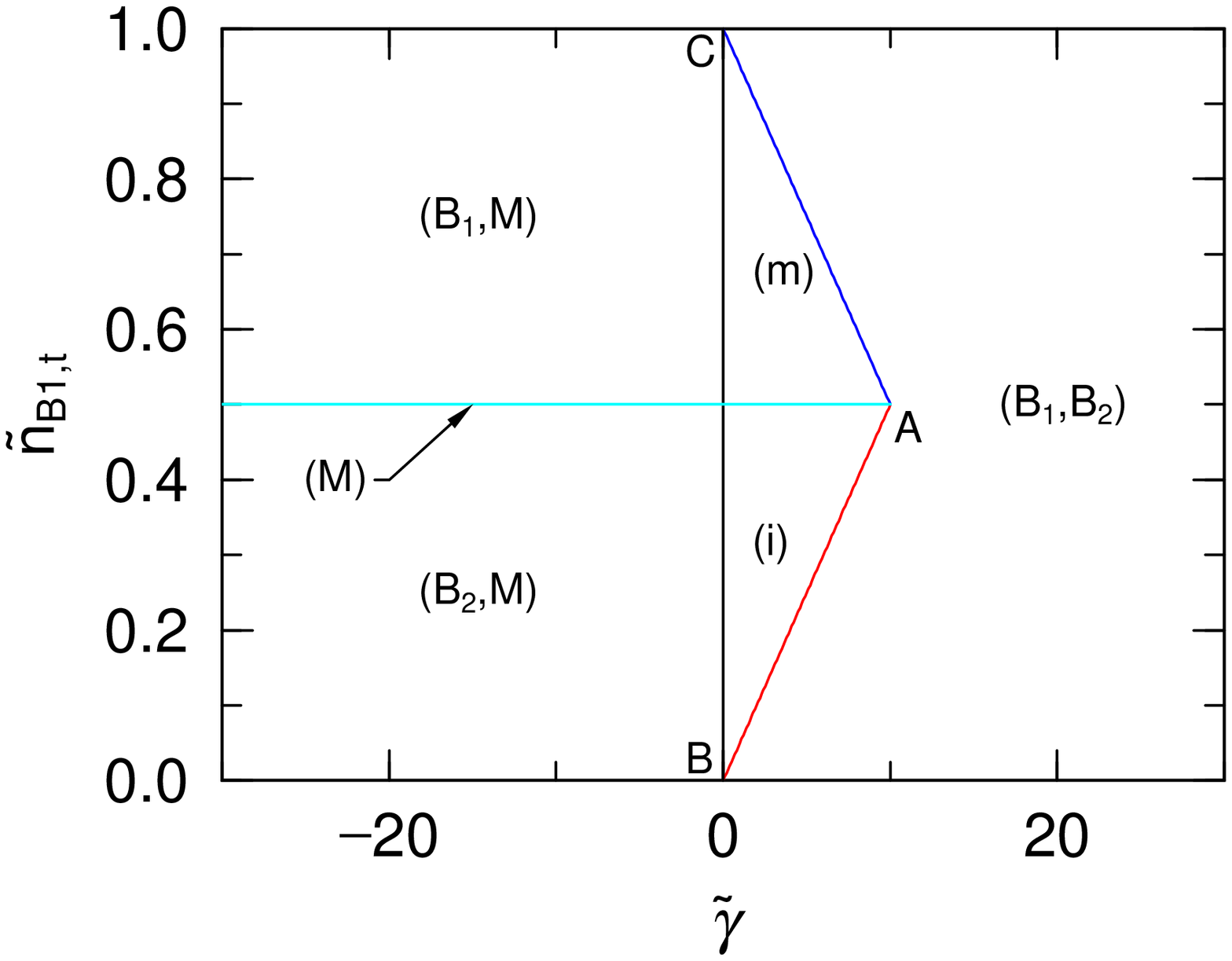} 
    \end{tabular}
  \end{center}
\caption{(Color online) 
The $T=0$ phase diagram of interacting BB mixtures 
with the same atom masses   
in the $\gamT$-$\nT_{B1,t}$ plane for $\gaT =20$ [(a), left] 
and $\gaT =-20$ [(b), right], 
where the phases $(i)$ and $(m)$ are coexisting phases of $(B1, B2)$, $(B2, M)$ 
and $(B1, B2)$, $(B1, M)$.}
\label{fig15}
\end{figure}

The $T=0$ phase diagrams 
of interacting $BB$ mixtures 
with $\mT_{B1} =\mT_{B2} =1/2$ 
are shown in Fig.~\ref{fig15}(a) ($\gaT=20$) and Fig.~\ref{fig15}(b) ($\gaT=-20$). 
In the case of $\gaT >0$, 
the phase diagrams have the mixed phases 
[the region $ABC$ in Fig.~\ref{fig15}(a)], 
which do not exist in the noninteracting cases. 
The position of the end point $A$ is given by (\ref{EqApB24})
in Appendix B:
\begin{equation}
     \gamT=-\frac{\gaT}{2}, 
\label{EqE26}
\end{equation}
which locates on the left side of the boundary $BC$ in Fig.~\ref{fig15}(a). 
We can understand that the phase structure in $\gaT>0$ is just PS1.
When $\gaT<0$,
Eq.~(\ref{EqE26}) shows that point $A$ moves 
across the boundary $BC$ and locates to the right of it; 
the mixed phase disappears and new phases occur
with coexisting locally stable equilibrium states 
 [Fig.~\ref{fig15}(b)]. 
That means that the phase structure is PS3 for $\gaT<0$.
As a result, we find $\gaT^{(BB)}_{c1} =\gaT^{(BB)}_{c2} =0$ 
and no PS2 phase structures exist in the interacting BB mixtures.

In the case of interacting FF mixtures, 
the critical values $\gaT^{(FF)}_{c1}$ and $\gaT^{(FF)}_{c2}$ 
become
\begin{eqnarray}
     \gaT^{(FF)}_{c1} &=& -\frac{2}{3} 
                           \left( \mT_{F1}^{-3/4} 
                                 +\mT_{F1}^{-3/4} \right)^{4/3},
\label{EqE27}\\
     \gaT^{(FF)}_{c2} &=& -2^{1/3} 
                           \left( \frac{3\pi^2}{\sqrt{2}} \right)^{2/3}
                           \frac{1}{\mT_{F1} \mT_{F2}},
\label{EqE28}
\end{eqnarray}
which become 
$\gaT^{(FF)}_{c1} \sim -25.5$ and 
$\gaT^{(FF)}_{c2} \sim -38.2$ 
in the case of $\mT_{F1} =\mT_{F2} =1/2$. 

\begin{figure}[b]
  \begin{center}
    \begin{tabular}{cc}
       \includegraphics[scale=0.4,angle=-90]{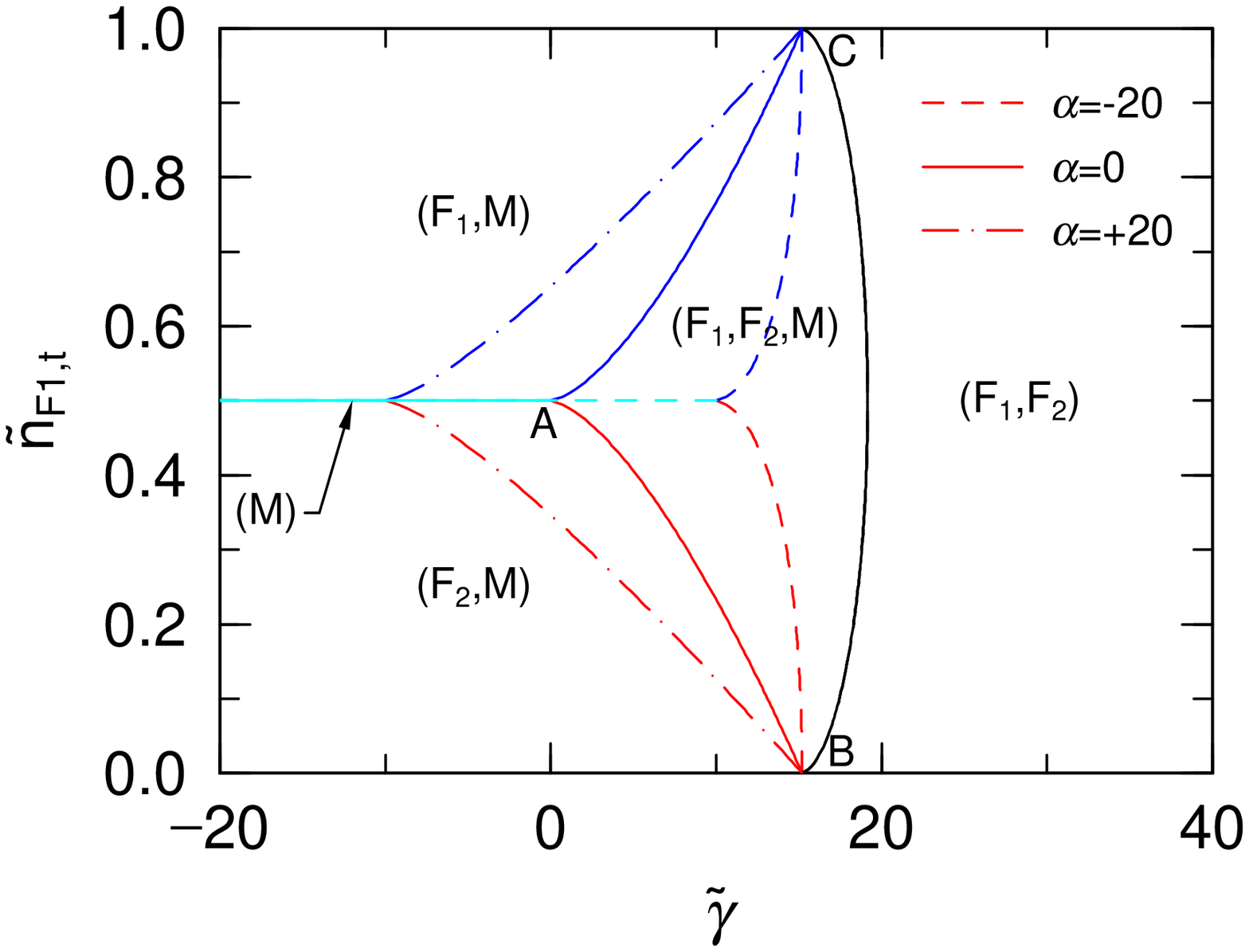} &
       \includegraphics[scale=0.4,angle=-90]{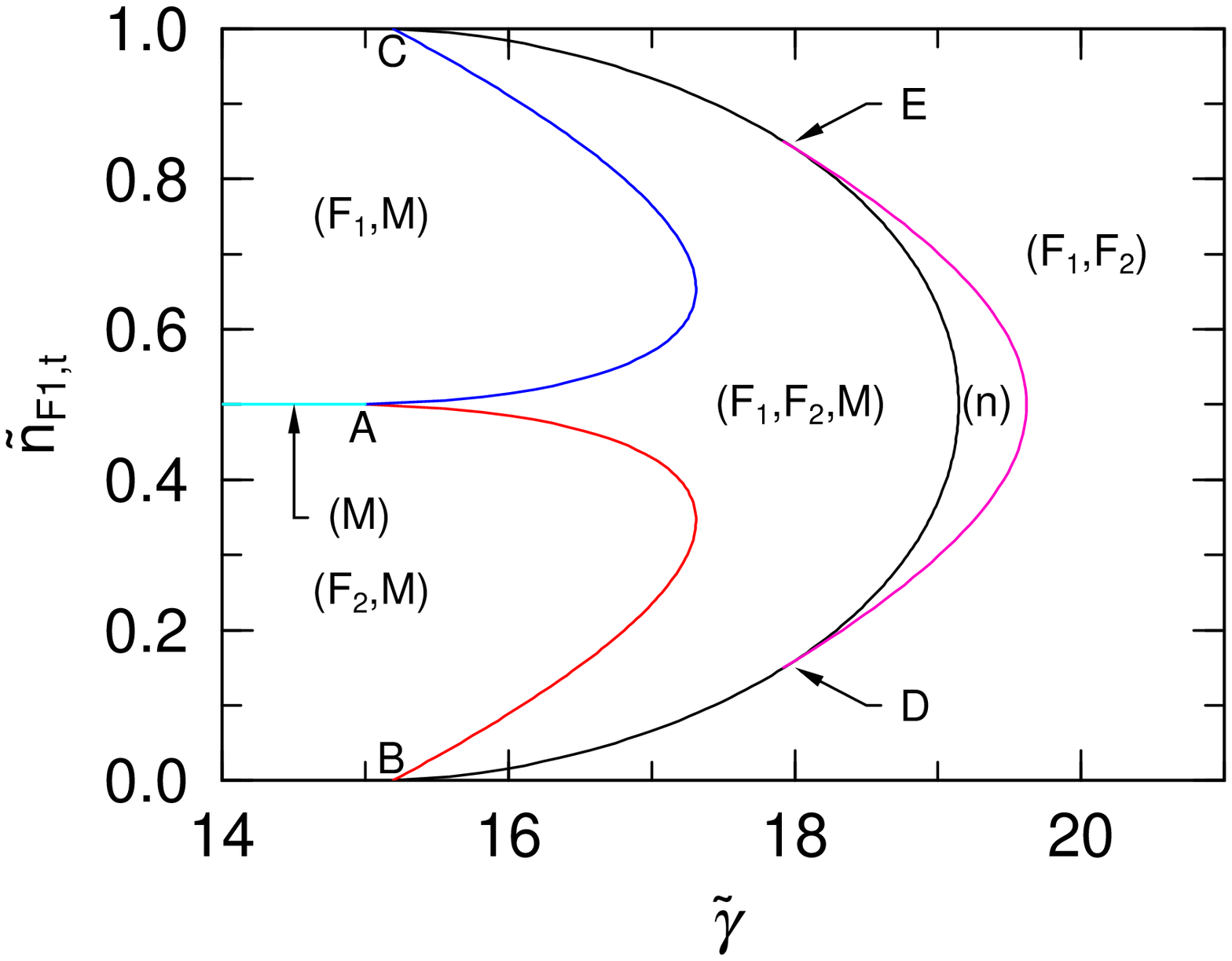} \\
       \includegraphics[scale=0.4,angle=-90]{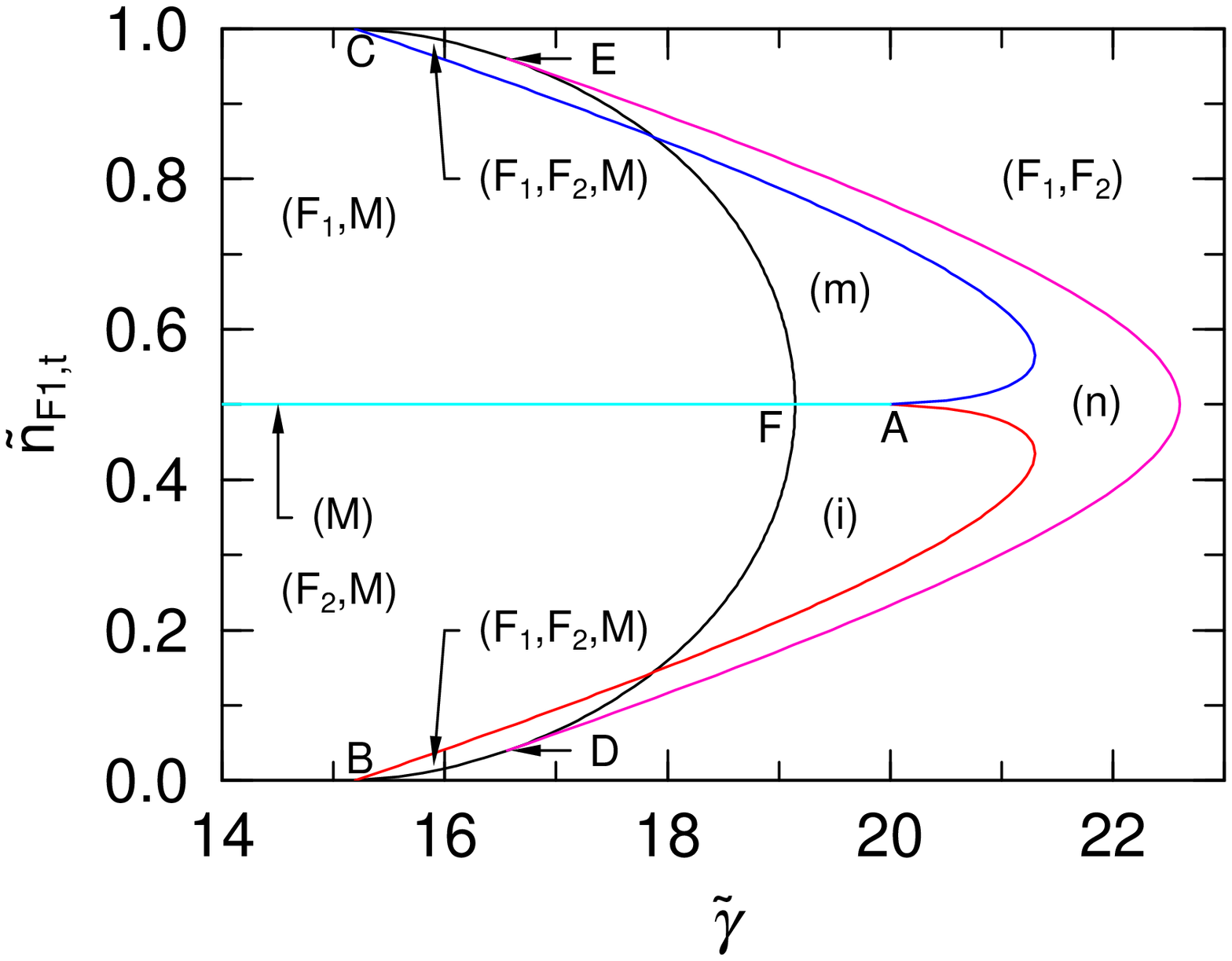} &
       \includegraphics[scale=0.4,angle=-90]{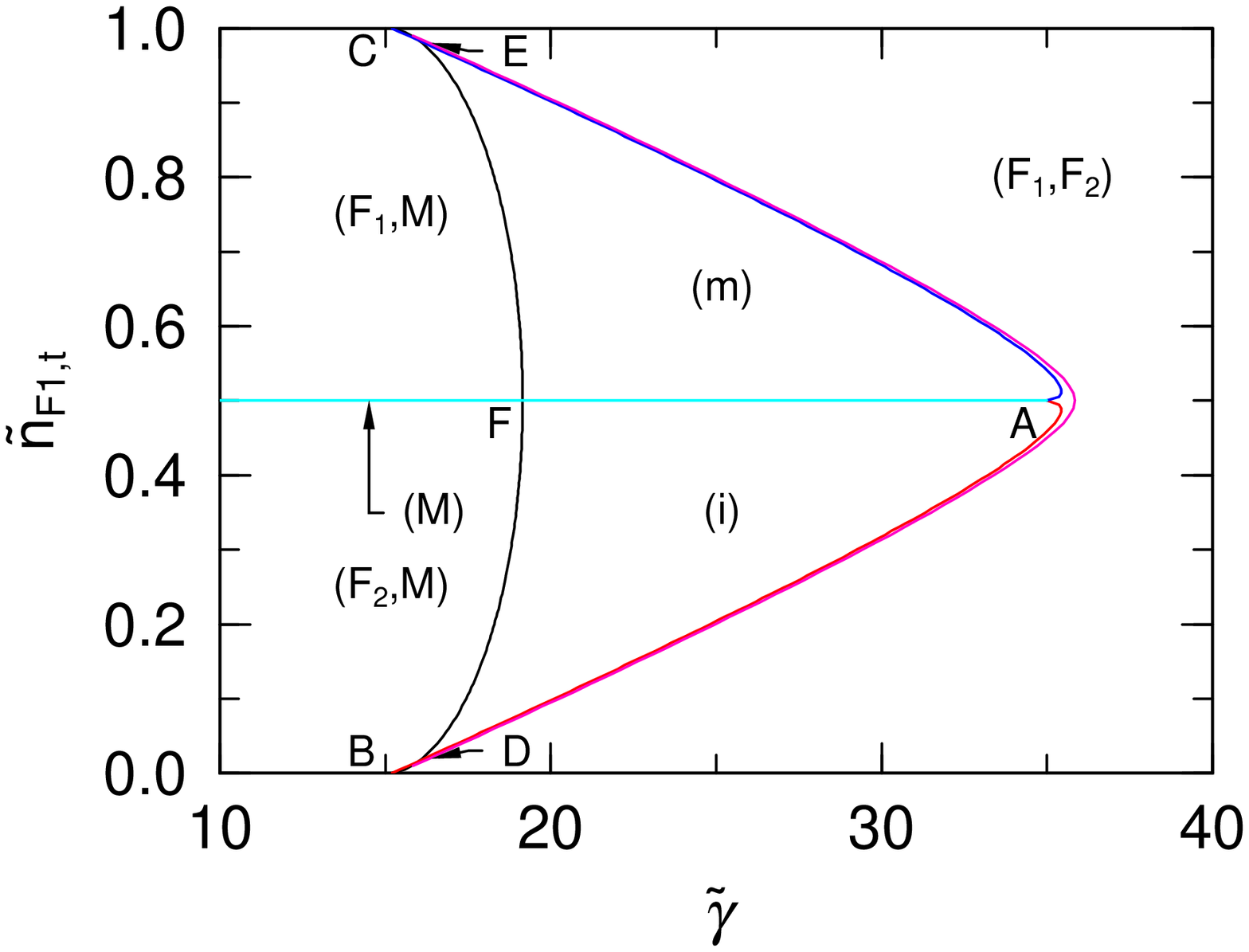} 
    \end{tabular}
  \end{center}
\caption{(Color online) 
The $T=0$ phase diagram of interacting FF mixtures 
with the same atom masses  
in the $\gamT$-$\nT_{F1,t}$ plane 
for $\gaT =20,0,-20$ [(a), top left], 
$-30$ [(b), top right],
$-40$ [(c), bottom left], 
and $-70$ [(d), bottom right], 
where the phases $(i)$, $(m)$, and $(n)$ are coexisting phases 
of $(F1, F2)$, $(F2, M)$; $(F1, F2)$, $(F1, M)$; and $(F1, F2)$, $(F1, F2, M)$.}
\label{fig16}
\end{figure}

In Fig.~\ref{fig16}, 
we show the $T=0$ phase diagrams 
for interacting FF mixtures 
with $\mT_{F1} =\mT_{F2} =1/2$. 
The change of the phase structure is essentially 
similar to that in the interacting BF mixtures.
In the PS1 structure [Fig.~\ref{fig16}(a)], 
the phase structures are obtained by 
deformation from the noninteracting ones.
The substructures with the coexisting states appear 
in the mixed phase in the PS2 structure [Fig.~\ref{fig16}(b)]. 
The end point $A$ in Fig.~\ref{fig16}(a) and \ref{fig16}(b) crosses the boundary $BC$ 
between the mixed and dissociated phases 
and new phases appear in the PS3 structure 
with the coexisting equilibrium states,
which have complex substructures [Fig.~\ref{fig16}(c)]; 
the structure becomes again simple for large negative values 
of $\gaT$ [Fig.~\ref{fig16}(d)].
%
%
\section{Summary and Outlook}
%

We have developed a quasichemical equilibrium theory 
for the molecular formation or dissociation processes 
in BF, FF, and BB mixtures of ultracold atomic gases 
and discussed atom-molecule equilibrium 
in these mixture. 

The law of mass actions has also been examined
for the mixtures; 
it is satisfied well at high $T$. 
We have shown that the quantum-statistical effects 
of the atoms and molecules, 
which become more effective in ultracold temperature,
give deviations from the law (law of quantum mass action). 
The quantum-statistical effects are shown to give different deviations 
for bosons and fermions, and,
in BF mixtures, both contributions have a tendency to 
cancel out at high $T$.

We have also discussed the effects of the interparticle interactions 
in the mixture
within the mean-field approximation at $T=0$ 
and evaluated the shifts of the $T=0$ phase structures 
of atom-molecule equilibrium in the mixtures.
Especially, in the case of large repulsive interactions
between atoms and molecules, 
the phase structures have been shown 
to change qualitatively 
with the occurrence of coexisting local-equilibrium states.
We have given the conditions for the coupling constants 
with which the phase-structure changes occur.
The atom-molecule equilibrium 
in interacting mixtures  
can be calculated also at finite temperatures 
within the present framework, 
and the results are planned to be published in another paper.

The other correlation effects beyond the mean-field approximations 
should also be important--for example, 
in the BCS-BEC crossover problem. 
Combining the method of the Beth-Uhlenbeck approach\cite{Beth,Schmidt}
with the present quasichemical equilibrium theory, 
we should discuss the correlation effects and the crossover problem 
from a less model-independent point of view. 
A study along these lines is now ongoing and will be presented in the near future.

The authors thank T.~Suzuki and T.~Takayama for many useful discussions. 
%
%
\appendix
%
%
\section{Asymptotic behaviors of Bose or Fermi functions \label{AppA}}
%

In this appendix, 
we derive some formulas 
of the asymptotic behaviors of the Bose and Fermi functions 
(\ref{EqB10}) and (\ref{EqB11})
at $\nu \sim 0$ and $\nu \sim \pm\infty$.

Before we discuss the asymptotic behavior, 
we prove the relation between $B_A$ and $F_A$:
\begin{equation}
     F_A(\nu) =B_A(\nu) -2^{1-A} B_A(2\nu).  \label{EqApA1}
\end{equation}
This formula is obtained by 
integrating both sides of the equation 
\begin{equation}
     \frac{x^{A-1}}{e^{x+\nu}+1} =\frac{x^{A-1}}{e^{x+\nu}-1} 
                           -\frac{2x^{A-a}}{e^{2x+2\nu}-1}. 
\label{EqApA2}
\end{equation}

Let us go to the asymptotic behavior of the functions $B_A(\nu)$ and $B_A(\nu)$
at $\nu \sim \infty$.
Using the expansion of the integrand of $B_A$,
\begin{equation}
     \frac{x^{A-1}}{e^{x+\nu}-1} 
          =\sum_{k=1}^\infty x^{A-1} e^{-k (x+\nu)},
\label{EqApA3}
\end{equation}
we obtain
\begin{equation}
     B_A(\nu) =\sum_{k=1}^\infty \frac{e^{-k\nu}}{\Gamma(A)}
                              \int_0^\infty x^{A-1} e^{-kx} dx
              =\sum_{k=1}^\infty \frac{e^{-k\nu}}{k^A},
\label{EqApA4}
\end{equation}
where we have used the formula of the gamma function:
\begin{equation}
     \Gamma(A) =\int_0^\infty z^{A-1} e^{-z} dz.
\label{EqApA5}
\end{equation}
Substituting (\ref{EqApA4}) into (\ref{EqApA1}), 
an expansion formula for $F_A(\nu)$ can be obtained:
\begin{equation}
     F_A(\nu) =\sum_{k=1}^\infty (-1)^{k-1} \frac{e^{-k\nu}}{k^A}.
\label{EqApA6}
\end{equation}
Taking the first terms in Eqs.~(\ref{EqApA4}) and (\ref{EqApA6}), 
we obtain the asymptotic behaviors of $B_A(\nu)$ and $F_A(\nu)$ 
at $\nu \sim \infty$:
\begin{equation}
     B_A(\nu) \sim F_A(\nu) \sim e^{-\nu}.  \label{EqApA7}
\end{equation}

\begin{figure}[t]
  \begin{center}
       \includegraphics[scale=0.5]{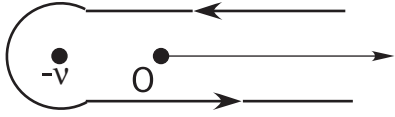}
  \end{center}
\caption{Hankel-type integral path in the complex-number plane.}
\label{figA1}
\end{figure}

We turn to the asymptotic behavior of $B_A(\nu)$ around $\nu =0$.
When $0 < \alpha <1$, 
the point $\nu=0$ becomes an irregular singular point, 
so that only the asymptotic expansion is obtained for $B_A$.
For this purpose, 
we consider the Hankel-type complex integral
\begin{equation}
     I =\frac{1}{\Gamma(A)} \int_{(\infty;-\nu+)}
        \frac{z^{A-1} dz}{e^{z+\nu}-1}. 
\label{EqApA8}
\end{equation}
The integration path $(\infty;-\nu+)$ and 
the cut line on the positive real axis are shown in Fig.~\ref{figA1}.
The phase branches of $z^{A-1}$ are fixed at $x^{A-1}$ 
or $e^{2\pi A i} x^{A-1}$ on the upper or lower parts of the cut line.
The integrand function has a pole at $z=-\nu$:
\begin{equation}
     \frac{z^{A-1}}{e^{z+\nu}-1} \sim \frac{(-\nu)^{A-1}}{z+\nu}.
\label{EqApA9}
\end{equation}

The integration path does not pass the singular point $z=-\nu$, 
so that we can evaluate the integral using the expansion (\ref{EqApA2}):
\begin{equation}
     I =\sum_{k=1}^\infty \frac{e^{-k\nu}}{\Gamma(A)} 
        \int_{(\infty;0+)} z^{A-1} e^{-kz} dz
       =\sum_{k=1}^\infty \frac{e^{-k\nu}}{k^A} 
        (e^{2\pi A i} -1),
\label{EqApA10}
\end{equation}
where we have used the Hankel-integral representation 
of the gamma function.

On the other hand, 
we can deform the integration path 
and divide it into the path $C_{-\nu}$ circulating 
around the singular point $z=-\nu$ 
and the Hankel-type path $(\infty;0+)$ around the origin $O$:
\begin{equation}
     I =\frac{1}{\Gamma(A)} 
        \int_{(\infty;-\nu+)}\frac{z^{A-1} dz}{e^{z+\nu}-1}
       =\frac{1}{\Gamma(A)} \left[ \oint_{C_{-\nu}} 
                                  +\int_{(\infty;0+)} \right]
                            \frac{z^{A-1} dz}{e^{z+\nu} -1}.
\label{EqApA11}
\end{equation}
The integral on $C_{-\nu}$ is evaluated by the theorem of residue, 
and that on $(\infty;0+)$ can be attributed to the real integral:
\begin{eqnarray}
     \frac{1}{\Gamma(A)} 
     \oint_{C_{-\nu}} \frac{z^{A-1} dz}{e^{z+\nu}-1}
          &=& \frac{2\pi i}{\Gamma(A)} (-\nu)^{A-1},  
     \label{EqApA12}\\
     \frac{1}{\Gamma(A)} 
     \int_{(\infty;0+)} \frac{z^{A-1} dz}{e^{z+\nu} -1}
          &=& \frac{e^{2\pi i A} -1}{\Gamma(A)} 
              \int_0^\infty \frac{x^{A-1} dx}{e^{x+\nu}-1} 
           =(e^{2\pi i A} -1) B_A(\nu).
     \label{EqApA13}
\end{eqnarray}
Combining Eqs.~(\ref{EqApA10}), (\ref{EqApA12}), and (\ref{EqApA13}), 
we obtain the asymptotic expansion of $B_A(\nu)$: 
\begin{equation}
     B_A(\nu) =\frac{\pi}{\sin\pi A} \frac{\nu^{A-1}}{\Gamma(A)} 
              +\sum_{k=1}^\infty \frac{e^{-k\nu}}{k^A}.
\label{EqApA14}
\end{equation}
Using the expansion of the series part in (\ref{EqApA14}),
\begin{equation}
     \sum_{k=1}^\infty \frac{e^{-k\nu}}{k^A} 
          =\sum_{k=1}^\infty \frac{1}{k^A} 
           \sum_{n=0}^\infty \frac{(-k\nu)^n}{n!}
          =\sum_{n=0}^\infty \frac{(-\nu)^n}{n!}
           \sum_{k=1}^\infty \frac{1}{k^{A-n}}
          =\sum_{n=0}^\infty (-1)^n \zeta(A-n) \frac{\nu^n}{n!},
\end{equation}
we obtain the power expansion formula by Opechowski\cite{Opechowski,terHaar}:
\begin{equation}
     B_A(\nu) =\frac{\pi}{\sin\pi A} \frac{\nu^{A-1}}{\Gamma(A)}
              +\sum_{n=0}^\infty (-1)^n \zeta(A-n) \frac{\nu^n}{n!}.
\label{EqApA15}
\end{equation}
Taking the leading term of the Opechowski formula, 
we obtain the asymptotic behavior of $B_A(\nu)$ around $\nu=0$:
\begin{equation}
     B_A(\nu) \sim \left\{
                   \begin{array}{@{\,}ll}
                   \zeta(A) & (A > 1), \\
                   -\ln{\nu}    & (A=1),   \\
                   \frac{\pi}{\sin\pi A} \frac{\nu^{A-1}}{\Gamma(A)}
                   +\zeta(A)     & (0 < A <1).
                   \end{array}\right.
     \label{EqApA16}
\end{equation}
Using Eq.~(\ref{EqApA1}),
the asymptotic formula of $F_A(\nu)$ around $\nu=0$ becomes
\begin{equation}
     F_A(\nu) \sim (1-2^{1-A}) \zeta(A).  \label{EqApA17}
\end{equation}
It should be noted that the residue term in $B_A(\nu)$ 
is canceled out in (\ref{EqApA17});
this is consistent with the fact that
no such singular terms exist in $F_A(\nu)$ originally.

The asymptotic formula of $F_A(\nu)$ at $\nu \sim -\infty$ 
is obtained by the Sommerfeld expansion formula\cite{terHaar,Sommerfeld}
\begin{equation}
     \int_0^\infty\frac{\phi'(u) du}{e^{u-\alpha}+1}
          \sim \phi(\alpha) 
              +\sum_{n=1}^\infty 2 F_{2n}(0) \phi^{(2n)}(\alpha),
\label{EqApA18}
\end{equation}
where $\phi(u)$ is a $\infty$-differentiable function 
and the coefficients $F_{2n}(0)$, the Fermi function at $\nu=0$, 
are represented by the Bernoulli numbers $B_n$:
\begin{equation}
     2 F_{2n}(0) =\frac{(1-2^{1-2n})(2\pi)^{2n}}{(2n)!} B_n.
\label{EqApA19} 
\end{equation}
Using Eq.~(\ref{EqApA18}), 
the asymptotic formula of $F_A(\nu)$ at $\nu \sim -\infty$ is obtained by
\begin{equation}
     F_A(\nu) \sim \frac{(-\nu)^A}{\Gamma(A+1)}.  \label{EqApA20}
\end{equation}

%
%
\section{Derivations of $T=0$ phase diagrams 
in BF and FF mixtures \label{AppB}}
%

\subsection{Noninteracting BF mixture}

Let us consider the mixed phase of the BF mixture
of atom masses $\mT_{B,F}$, 
which corresponds to the central region 
in the phase diagram (Fig.~\ref{fig1}). 
The equilibrium condition (\ref{EqB17}) becomes 
\begin{equation}
     0 +\left( \frac{3\pi^2}{\sqrt{2}} \right)^{2/3}
                       \frac{1}{\mT_F} (\nT_F)^{2/3}  
       -\left( \frac{3\pi^2}{\sqrt{2}} \right)^{2/3}
                        (\nT_M)^{2/3}
       =\Delta\ET_M,
\label{EqApB1}
\end{equation}
where we have used $\mu_B=0$ (BEC state) 
and the Fermi energy formula (\ref{EqC4}) for $F$ 
and $M$ with $\mT_M \sim 1$.

The boundaries $AB$, $BC$, and $CA$ in Fig.~\ref{fig1} 
are obtained by setting $\nT_B =0$, $\nT_M =0$, and $\nT_F =0$, 
respectively:
\begin{eqnarray}
     &AB:\qquad& 
          \Delta\ET_M =\frac{1}{2\mT_F} [ 6\pi^2 (1-2\nT_{B,t}) ]^{2/3} 
                      -\frac{1}{2} [ 6\pi^2 \nT_{B,t} ]^{2/3}, 
\label{EqApB2} \\ 
     &BC:\qquad& 
          \Delta\ET_M =\frac{1}{2\mT_F} [ 6\pi^2 \nT_{F,t} ]^{2/3},
\label{EqApB3} \\ 
     &CA:\qquad& 
          \Delta\ET_M =-\frac{1}{2} [6\pi^2 \nT_{F,t} ]^{2/3}.
\label{EqApB4} 
\end{eqnarray}
and the end points of the boundaries, $A$, $B$, and $C$,
are obtained 
from (\ref{EqApB4}) with $\nT_{F,t}=1/2$, 
(\ref{EqApB3}) with $\nT_{F,t}=1$, and 
(\ref{EqApB3}) with $\nT_{F,t}=0$:
\begin{eqnarray}
     &A:\qquad& 
          \Delta\ET_M =-\frac{(3\pi^2)^{2/3}}{2} \sim -4.8, 
\label{EqApB5} \\ 
     &B:\qquad& 
          \Delta\ET_M =\frac{1}{2\mT_F} (6\pi^2)^{2/3} \sim \frac{7.6}{\mT_F},
\label{EqApB6} \\ 
     &C:\qquad& 
          \Delta\ET_M =0.
\label{EqApB7} 
\end{eqnarray}
The boundaries and end points in Fig.~\ref{fig1} are obtained 
from the above formulas for $\mT_F =1/2$.

\subsection{Noninteracting FF mixture}

In the case of the phase diagram of the FF mixture (Fig.~\ref{fig4}) 
with masses $\mT_{F1}$ and $\mT_{F2}$, 
the equilibrium condition (\ref{EqB17}) 
in the mixed phase becomes
\begin{equation}
     \left( \frac{3\pi^2}{\sqrt{2}} \right)^{2/3}
                      \frac{1}{\mT_{F1}} (\nT_{F1})^{2/3}
     +\left( \frac{3\pi^2}{\sqrt{2}} \right)^{2/3}
                      \frac{1}{\mT_{F2}} (\nT_{F2})^{2/3}  
       =\Delta\ET_M,
\label{EqApB8}
\end{equation}
where we have used $\mu_M=0$ (BEC state) 
and the Fermi energy formula (\ref{EqC4}) for $F{1}$ and $F{2}$..

The boundaries $AB$, $BC$, and $CA$ in Fig.~\ref{fig4}
are obtained by setting $\nT_{F2} =0$, $\nT_M =0$, and $\nT_{F1} =0$, 
respectively:
\begin{eqnarray}
     &AB:\qquad& 
          \Delta\ET_M =\frac{1}{2\mT_{F1}} [ 6\pi^2 (1-2\nT_{F1,t}) ]^{2/3}, 
\label{EqApB9} \\ 
     &BC:\qquad& 
          \Delta\ET_M =\frac{1}{2\mT_{F1}} [ 6\pi^2 \nT_{F1,t} ]^{2/3}
                      +\frac{1}{2\mT_{F2}} [ 6\pi^2 \nT_{F2,t} ]^{2/3},
\label{EqApB10} \\ 
     &CA:\qquad& 
          \Delta\ET_M =\frac{1}{2\mT_{F2}} [ 6\pi^2 (2\nT_{F1,t}-1) ]^{2/3},
\label{EqApB11} 
\end{eqnarray}
and the points $A$, $B$ and $C$ are obtained 
from (\ref{EqApB9}) with $\nT_{F1,t}=1/2$, 
(\ref{EqApB9}) with $\nT_{F1,t}=0$, and 
(\ref{EqApB11}) with $\nT_{F1,t}=1$:
\begin{eqnarray}
     &A:\qquad& 
          \Delta\ET_M =0,
\label{EqApB12} \\ 
     &B:\qquad& 
          \Delta\ET_M =\frac{1}{2\mT_{F1}} (6\pi^2)^{2/3}
                      \sim \frac{7.6}{\mT_{F2}},
\label{EqApB13} \\ 
     &C:\qquad& 
          \Delta\ET_M =\frac{1}{2\mT_{F2}} (6\pi^2)^{2/3}.
                      \sim \frac{7.6}{\mT_{F1}}.
\label{EqApB14} 
\end{eqnarray}
The boundaries and end points in Fig.~\ref{fig4} are obtained 
from the above formulas with $\mT_{F1} =\mT_{F2} =1/2$.

\subsection{Interacting BF mixture}

The $T=0$ phase diagram
of the interacting BF mixture 
with masses $\mT_{B,F}$ 
is obtained by the equilibrium condition (\ref{EqE22}) 
for $T=0$: 
\begin{equation}
     0 +\left( \frac{3\pi^2}{\sqrt{2}} \right)^{2/3}
                       \frac{1}{\mT_F} (\nT_F)^{2/3}  
       -\left( \frac{3\pi^2}{\sqrt{2}} \right)^{2/3}
                        (\nT_M)^{2/3}
       =\gaT \nT_M +\gamT.
\label{EqApB15}
\end{equation}
In the case of $\gaT > \gaT^{(BF)}_{c1}$, 
the boundaries and end points of the phases 
can be obtained in the same manner as 
those in noninteracting cases.

The results are
\begin{eqnarray}
     &AB:\qquad& 
          \gamT =\left( \frac{3\pi^2}{\sqrt{2}} \right)^{2/3} 
                  \left[\frac{1}{\mT_F} (\nT_{F,t} -\nT_{B,t})^{2/3} 
                     -\nT_{B,t}^{2/3} \right]
                  -\gaT \nT_{B,t}, 
\label{EqApB15a} \\ 
     &BC:\qquad& 
          \gamT =\left( \frac{3\pi^2}{\sqrt{2}} \right)^{2/3} 
                  \frac{1}{\mT_F} \nT_{F,t}^{2/3},
\label{EqApB16} \\ 
     &CA:\qquad& 
          \gamT =-\left( \frac{3\pi^2}{\sqrt{2}} \right)^{2/3} 
                   \nT_{F,t}^{2/3}
                  -\gaT \nT_{F,t},
\label{EqApB17} 
\end{eqnarray}
for the phase boundaries 
($\nT_{B,t} +\nT_{F,t} =1$), 
and
\begin{eqnarray}
     &A:\qquad& 
          \gamT=-\left( \frac{3\pi^2}{2\sqrt{2}} \right)^{2/3}
                  -\frac{\gaT}{2}
                 \sim -4.78 -\frac{\gaT}{2},
\label{EqApB18} \\ 
     &B:\qquad& 
          \gamT =\frac{1}{\mT_F} \left( \frac{3\pi^2}{\sqrt{2}} \right)^{2/3}
                 \sim \frac{7.6}{\mT_F},
\label{EqApB19} \\ 
     &C:\qquad& 
          \gamT =0.
\label{EqApB20} 
\end{eqnarray}
The boundaries and end points in Fig.~\ref{fig10} are obtained 
from the above formulas with $\mT_F =\mT_B =1/2$.

\subsection{Interacting BB mixture}

The $T=0$ phase diagram
of the interacting BB mixture 
with masses $\mT_{B1,B2}$ 
is obtained by the equilibrium condition (\ref{EqE22}).

At $T=0$, all chemical potentials appearing in (\ref{EqE22}), 
which are bosonic, 
can take two alternative possibilities: 
$\muT'_k =0$ with $\nT_k \neq 0$ (BEC) 
or $\muT'_k<0$ with $\nT_k =0$ ($k=B1,B2,M$).

In the mixed phase, 
the BEC conditions $\muT'_{B1}=\muT'_{B2}=\muT'_M=0$ 
give $\gaT \nT_M +\gamT =0$; 
the boundaries, $AB$, $BC$, and $CA$, 
are obtained by substituting 
$\nT_{B2} =0$, $\nT_M =0$, and $\nT_{B1} =0$, respectively:
\begin{eqnarray}
     &AB:\qquad& 
          \gamT =-\gaT \nT_{B1,t}, 
\label{EqApB21} \\ 
     &BC:\qquad& 
          \gamT =0,
\label{EqApB22} \\ 
     &CA:\qquad& 
          \gamT =-\gaT (1-\nT_{B1,t}),
\label{EqApB23} 
\end{eqnarray}
where we have used the constraints 
$\nT_{B1} +\nT_M =\nT_{B1,t}$ and 
$\nT_{B2} +\nT_M =\nT_{B2,t}$.
The end points are obtained by
\begin{eqnarray}
     &A:\qquad& 
          \gamT=-\frac{\gaT}{2}, 
\label{EqApB24} \\ 
     &B:\qquad& 
          \gamT =0,
\label{EqApB25} \\ 
     &C:\qquad& 
          \gamT =0.
\label{EqApB26} 
\end{eqnarray}
It should be noted that the boundaries and the end points 
are independent of the atom masses $\mT_{B1,B2}$.

\subsection{Interacting FF mixture}

The $T=0$ phase diagram
of the interacting FF mixture 
with masses $\mT_{B,F}$ 
is obtained by the equilibrium condition (\ref{EqE22}). 
for $T=0$: 
\begin{equation}
     \left( \frac{3\pi^2}{\sqrt{2}} \right)^{2/3}
                      \frac{1}{\mT_{F1}} (\nT_{F1})^{2/3}
     +\left( \frac{3\pi^2}{\sqrt{2}} \right)^{2/3}
                      \frac{1}{\mT_{F2}} (\nT_{F2})^{2/3}  
       =\gaT \nT_M +\gamT.
\label{EqApB27}
\end{equation}
In the case of $\gaT > \gaT^{(FF)}_{c1}$, 
the boundaries and end points of the phases 
can be obtained in the same manner as 
those in noninteracting cases.

The results are
\begin{eqnarray}
     &AB:\qquad& 
          \gamT =\frac{1}{2\mT_{F1}} [ 6\pi^2 (1-2\nT_{F1,t}) ]^{2/3}
                -\gaT \nT_{F1,t}, 
\label{EqApB28} \\ 
     &BC:\qquad& 
          \gamT =\frac{1}{2\mT_{F1}} [ 6\pi^2 \nT_{F1,t} ]^{2/3}
                      +\frac{1}{2\mT_{F2}} [ 6\pi^2 \nT_{F2,t} ]^{2/3},
\label{EqApB29} \\ 
     &CA:\qquad& 
          \gamT =\frac{1}{2\mT_{F2}} [ 6\pi^2 (2\nT_{F1,t}-1) ]^{2/3}
                +\gaT (\nT_{F1.t} -1),
\label{EqApB30} 
\end{eqnarray}
for the phase boundaries 
($\nT_{B,t} +\nT_{F,t} =1$), 
and
\begin{eqnarray}
     &A:\qquad& 
          \gamT =-\frac{\gaT}{2},
\label{EqApB31} \\ 
     &B:\qquad& 
          \gamT =\frac{1}{2\mT_{F1}} (6\pi^2)^{2/3}
                      \sim \frac{7.6}{\mT_{F2}},
\label{EqApB32} \\ 
     &C:\qquad& 
          \gamT =\frac{1}{2\mT_{F2}} (6\pi^2)^{2/3}.
                      \sim \frac{7.6}{\mT_{F1}}.
\label{EqApB33} 
\end{eqnarray}
The boundaries and end points in Fig.~\ref{fig14} are obtained 
from the above formulas with $\mT_{F1} =\mT_{F2} =1/2$.

%
\end{document}